# Selection of multiple donor gauges

# via Graphical Lasso for estimation of daily streamflow time series

**German A. Villalba[1], Xu Liang[1], and Yao Liang[2]**

[1] Department of Civil and Environmental Eng., University of Pittsburgh, Pittsburgh, PA, USA.

[2] Department of Computer and Information Science, Indiana Univ.-Purdue Univ. Indianapolis, IN, USA.

Corresponding author: Xu Liang ([xuliang@pitt.edu](xuliang@pitt.edu))

**Key Points:**

- A novel algorithm is presented for selecting multiple donor gauges based on graphical Markov modeling for streamflow network via Glasso.

- A general method is presented using a multiple objective optimization for systematic donor gauge selection with noisy data.

- The new method shows superior results comparing to available methods in different applications.

**Abstract**

A fundamental challenge in estimations of daily streamflow time series at sites with incomplete records is how to *effectively* and *efficiently* select reference/donor gauges from an existing gauge network to infer the missing data. While research on estimating missing streamflow time series is not new, the existing approaches either use a single reference streamflow gauge or employ a set of 'ad-hoc' reference gauges, leaving a systematic selection of reference gauges as a long-standing open question. In this work, a novel method is introduced that facilitates systematical selection of multiple reference gauges from any given streamflow network. The idea is to mathematically characterize the network-wise correlation structure of a streamflow network via graphical Markov modeling, and further transforms a dense network into a sparsely connected one. The resulted underlying sparse graph from the graphical model encodes conditional independence conditions among all reference gauges from the streamflow network, allowing determination of an optimum subset of the donor gauges. The sparsity is discovered by using the Graphical Lasso algorithm with an L1-norm regularization parameter and a thresholding parameter. These two parameters are determined by a multi-objective optimization process. Furthermore, the graphical modeling approach is employed to solve another open problem in gauge removal planning decision (e.g., due to operation budget constraints): which gauges to remove would statistically guarantee the least loss of information by estimations from the remaining gauges? Our graphical model-based method is demonstrated with daily streamflow data from a network of 34 gauges over the Ohio River basin.

## 1. Introduction

Continuous daily streamflow time series are important for a wide variety of applications in hydrology and water resources. Such applications include water supply management, hydropower development, flood and drought control, forecasting of agricultural yield, ecological flow assessment, navigation, rainfall runoff model calibration, design of engineering structures such as highways and reservoirs, and many others (e.g., Archfield & Vogel, 2010; Farmer & Vogel, 2013; Parada & Liang, 2010; Razavi et al., 2013; Shu & Ouarda, 2012; Stagnitta et al., 2018). However, continuous streamflow data are not available oftentimes where needed due to data gaps in the recorded time series at gauged stations (e.g., Huang & Yang, 1998). Also, data gaps of different time periods could exist at different gauged locations within a large river basin (e.g., Hughes & Smakhtin, 1996). Furthermore, there is an increasing decline, such as the shutting-down of gauged stations, in the hydrometric network density worldwide (Mishra & Coulibaly, 2009; Samuel et al., 2011; Hannah et al., 2011). For example, the U.S. Geological Survey (USGS) is in the process of discontinuing operations of some streamflow stations nationwide due to budget cuts (USGS, 2019) which has been a serious concern (Lanfear & Hirsch, 1999; Stokstad, 2001; Vorosmarty et al., 2001; Witze, 2013). In fact, numerous streamflow stations with long historical records have been discontinued in the past (e.g., Hannah et al., 2011; Mishra & Coulibaly, 2009). For example, Hannah et al. (2011) pointed out that there is shrinkage in river gauging networks worldwide. The World Water Assessment Programme (2009) indicated that from 1980 to 2004, 2051 river gauges in the USA that have data records of more than 30 years were discontinued, leaving only 7360 stations by 2005. Another example is that there were a total number of 467 streamflow stations with a record of more than 30 years being closed in Canada from 1987 to 2007 (Hannah et al., 2011).  there is, therefore, a critical need to develop an effective, objective and general method to fill in data gaps and extend data records for those gauges that have been or will be discontinued.

The estimation techniques for continuous daily streamflow time series can be classified into two broad categories: (1) hydrologic model–dependent methods (e.g., Fernandez et al., 2000; Parada & Liang, 2010) and (2) hydrologic model–independent methods or data-driven methods (Razavi et al., 2013), which are also called statistical methods (Loukas & Vasiliades, 2014) or hydrostatistical methods (Farmer & Vogel, 2013). Work related to the former category is abundant but has its limitations, such as determining reasonable model parameters (e.g., Farmer & Vogel, 2013; He et al., 2011; Razavi et al., 2013; Zhang & Chiew, 2009). In this study, we focus on the work related to the latter category. Techniques and researches developed in this category include methods that require donor stations and those, such as regionally-based methods, that do not require a donor station. Studies have shown that methods employing donor stations significantly outperform those of the regionally-based methods (e.g. Stagnitta et al., 2018; Swain & Patra, 2017). The donor-based methods in use include, regression-based such as the MOVE (maintenance of variance) method (e.g., Helsel and Hirsch, 2002; Hirsch, 1982), scaling-based (e.g., Laaha & Bloschl, 2005; Stagnitta et al., 2018), and machine learning-based methods (e.g., Dibike & Solomatine, 2001; Solomatine & Ostfeld, 2008). Farmer and Vogel (2013) summarized the general procedure of them as a two-step process: Step 1: selection of one or multiple donor gauges; and Step 2: inference of the streamflow time series at the target site (e.g., partially gauged/incomplete) based on information from the donor gauge(s).

At present, the widely used techniques in the donor gauge selection include nearest neighbour method, maximum correlation method, minimum variance method, and methods related to an assessment of the hydrologic similarity, such as those based on similar drainage area, similar annual precipitation, or other similar basin characteristics, between the target and the donor catchments (e.g., Archfield & Vogel, 2010; Arsenault & Brissette, 2014; Farmer & Vogel, 2013; Halverson & Fleming, 2015; Mishra & Coulibaly, 2009; Smakhtin, 1999; Smakhtin et al., 1997). The similar drainage area method is effective if the climate and hydrologic regimes at the target and donor sites are similar and that the area is the only dominant factor affecting the streamflow. However, such requirements are generally not met, because a number of factors can significantly change the scaling relationship, such as, the orographic effects where the site at a different elevation is likely to receive a different amount of rainfall and thus a different amount of runoff per unit area; a site in the windward side of the mountain versus the site in the rain shadow side where the rainfall characteristics are dramatically different; differences in slopes, soil types, land cover and land use which can affect the conditions of runoff generation, leading to differences in the basin's response to rainfall; and differences in temperature that affect the evapotranspiration losses and runoff per unit area. Methods based on the similarity concept with multiple comprhensive basin characteristics are subjective and mixed results have been reported (e.g., Sellami et al., 2014; Stagnitta et al., 2018; Swain & Patra, 2017). Some methods originally developed for designing hydrometric networks could also be, in principle, useful for donor gauge selection or gauge removal. These methods include clustering algorithms and the use of entropy, both of which are devised for identifying redundant stations in a network. Among these two, the former identifies redundant stations when they are clustered within the same group for rationalizing a hydrometric network (Burn & Goulter, 1991; Mishra & Coulibaly, 2009); the latter looks into the information content each individual site contains, and gauges containing redundant information may be removed from the network. With further development, these methods could potentially also be useful in donor gauge selection applications. However, with these existing methods for donor gauge selection, long-standing open questions remain: what constitutes an appropriate number of donor gauges to be selected; and how to select them in a systematic and consistent way. In this study, we propose a novel approach to tackle these open questions.

Among the aforementioned existing donor selection methods, the nearest gauge method is a convenient and widely used one due to its simplicity and minimum data requirements (e.g., Asquith et al., 2006; Emerson et al., 2005; Farmer & Vogel, 2013; Mohamoud & M., 2008; Stagnitta et al., 2018). For example, Farmer & Vogel, (2013) adopted this simple distance-based method (referred to as Dist hereafter) in the donor selection procedure. Stagnitta et al. (2018) showed that the nearest gauge method led to better inferred low streamflow in most of the cases compared to other donor selection methods including the most similar drainage area, site producing the minimum variance flow estimator within 100 km from the target site, and a combination of the similarity in drainage area and the estimated flow variance, where the same inference methods were employed. The correlation-based approach (referred to as Corr hereafter), another popular method, uses the pair-wise correlation between the target and each of the donor gauges. It has been found that Corr is generally better than Dist (e.g., Archfield and Vogel, 2010; Ergen & Kentel, 2016; He et al., 2011) because information contained in the streamflow data is more effectively used in Corr and the marginal independence between any pair of gauges can be easily determined (Koller & Friedman, 2009). Two gauges are deemed independent of each other if their pair-wise correlation is below a given threshold (Halverson & Fleming, 2015).

Although using a single donor gauge to estimate streamflow time series has been a dominant approach (Farmer, 2016) based on either distance or correlation, Smakhtin et al. (1997) and Smakhtin (1999) proposed to use more than one donor gauge from nearby gauges to improve the streamflow estimations for the target basins. Stagnitta et al. (2018) also compared a single donor gauge versus multiple donor gauges in which the multiple donor gauges were selected based on criteria such as gauges within a given distance (e.g., 100 km and 200 km) and gauges having drainage areas within ±50%. Their findings for the low flow inferences were that the single donor gauge based on the nearest neighbor method is the best most of the time, while the second best is to use multiple donor gauges selected based on the distance within 100 km. Granato (2009) summarized that in general, the use of multiple donor gauges provides more accurate estimates of streamflow for gap-filling and/or record extension. Harvey et al. (2012) compared 15 different existing methods, including least-squares linear regression, least-squares multiple linear regression, MOVE.1, catchment area scaling, and long-term mean scaling, for gap-filling for a single donor and two donor gauge cases at 26 streamflow stations in the United Kingdom (UK), and concluded that overall, methods using two donor gauges outperformed their counterparts that used a single donor gauge. Also, the catchment area scaling method provided the poorest performance. The challenges of these approaches include: (1) how to measure the similarity; (2) how to systematically determine which gauges should be the donor gauges; and (3) how many donor gauges each individual target gauge should have. As can be seen, these multiple donor gauge selection criteria are basically ad hoc and some aspects of them are highly subjective.

In addition to the single or multiple donor gauge selection approaches, there are some alternative methods which try to use all of the available gauges in the study region. For example, researchers (e.g., Farmer, 2016; Skøien & Blöschl, 2007; Solow & Gorelick, 1986) applied geostatistical methods, such as methods from the kriging family, as effective alternatives. The kriging method is an spatial interpolation technique that estimates values at target locations as a linear weighted combination of the observations from different locations. The weights are assigned based on a variogram model which is usually fitted based on the variance between observations as a function of the distance between locations. The kriging method thus avoids the problem of donor gauge selection by using all available gauges in a region. While the kriging method is useful in some situations (Villeneuve et al., 1979), Virdee & Kottegoda (1984) noticed that a major problem with kriging is the lack of data with needed density. Thus, for a commonly encountered situation in which the density of streamflow network is sparse, kriging is not a good candidate. Along the same line of the kriging method, Archfield and Vogel (2010) developed a procedure called Map correlation method that uses time series from streamflow gauges in the study area to create a correlation map based on a kriging method and then uses that map to estimate the correlation between a given target location and nearby gauges. They concluded that the accuracy based on the most correlated gauge outperforms the one based on the distance in most cases.

Halverson & Fleming (2015) were one of the first groups who used network theory to investigate properties of the streamflow network consisting of 127 gauges located in the Coast Mountains of British Columbia and Yukon in Canada. Based on 10 years of daily streamflow data, the study constructed a streamflow network by connecting streamflow gauges if the pair-wise correlation between two gauges is or greater than 0.7. They then further applied the edge betweenness algorithm to the streamflow network to investigate its community (i.e., cluster) structure so that gauges with similar hydrologic behaviors were grouped into the same community. They concluded, based on the identified communities, that there are three types of gauges within the network that are important and should not be removed if the network is under a budget constraint. These three types of gauges are: (1) the gauges within each community that have a large number of intracommunity links, (2) gauges with high betweenness values which embed information about multiple communities, and (3) gauges that are members of single-membership or small-membership communities. While Halverson & Fleming (2015) studied which gauges in a streamflow network could be removable via graph theory-based community detection algorithm, their work, due to its different focus, did not address how to select donor gauges for each individual target gauge to infer the missing information after these target gauges are removed. Neither was the issue of high noises in the data that affected the correlation structure addressed. Also, among the removable gauges, their approach does not offer any quantitative estimates on the differences of the information loss for each individual gauge's removal, and thus does not provide any possible preferred order in the gauge removal process.

By and large, donor gauge selection is a long-standing challenge for streamflow estimation. It appears that the distance-based and correlation-based single donor methods (i.e., pair-wise marginal independence approach) are less subjective and generate more or less consistent results but there is plenty of room for significant improvement of the accuracy of their estimations. On the other hand, when multi-donor gauge selection methods are applied, due to their subjective and 'ad hoc' selection process on the donor gauges (including the repeated use of distance- or correlation- based single donor method), they are in general subjected to significant inconsistency in their performance. The pair-wise correlation approach represents a localized view and misses the global dependence structure of the daily streamflow embodied by its underlying streamflow network. This is so because the conditional independencies among gauges in a streamflow network are typically not apparent in the correlation matrix but in its inverse matrix, i.e., the precision matrix (Koller & Friedman, 2009), the existing pair-wise correlation-based methods on multi-donor selection process are thus ineffective.

It is critical to simultaneouly achieve objectivity and consistently good performance in donor gauge selection. Our approach, as presented in this paper, is to embrace global dependence structure, which makes it possible to systematically tackle the long-standing challenging donor gauge selection problem. In the process, a novel method is developed that can explictly and effectively consider the global dependence structure among a set of gauge stations from the viewpoint of the entire gauge network based on conditional independence conditions embedded in the graphical Markov models. We devise a new algorithm that makes use of such identified dependence structure to systematically optimize multiple donor gauges selection. In contrast to the commonly used correlation matrix, our method extracts and expresses the hidden information of conditional independence structure of the underlying streamflow network using a sparse precision matrix. This formulation further provides a framework that could be used in better understanding the essence of how multiple donor (or reference) gauge selection decision is made. Furthermore, we devise an additional new algorithm based on our new donor gauge selection method to determine which gauge(s), when required, could be discontinued from an existing streamflow network with the least loss of information.

The remainder of this paper is organized as follows. Section 2 briefly describes the necessary mathematical background. Section 3 presents our new approach. Section 4 presents two applications with our new approach: one is to fill data gaps for active gauges or to extend data records for terminated gauges or inactive gauges which are defined as those that are no longer collecting data but collected data in the past; and the other is to decide which gauges are best candidates to be removed from an existing hydrometric network with the least loss of information. Section 5 describes data used for evaluating the new approach and algorithms presented in Sections 3 and 4. Section 6 presents results and discussions. Finally, Section 7 provides a summary of the main findings from this work.

## 2.  Mathematical Background

To make this paper self-contained, some mathematical background is briefly reviewed in this section. Section 2.1 describes streamflow estimations with linear regression where a single, multiple or all available gauges can be used as donor gauges. Section 2.2 describes covariance and precision matrices. Section 2.3 reviews Gaussian graphical models which serves as the mathematical foundation of our donor gauge selection work. Section 2.4 describes the Graphical Lasso algorithm which makes precision matrix sparse.

### 2.1 Streamflow estimation with linear regression

2.1.1 A single donor gauge

Let $\mathbf{Q_j}$ and $\mathbf{Q_i}$ represent the streamflow vector with $n$ records from a target and a donor gauge, respectively, and assume that the estimated streamflow time series, i.e., a vector with $n$ records as well, at the target location ($\hat{\mathbf{Q}}_\mathbf{j}$) is obtained by transferring the streamflow time series from a single donor gauged catchment by a scaling function such that $\widehat{\mathbf{Q}}_\mathbf{j}$ is an approximation of $\mathbf{Q_j}$ (e.g., Archfield & Vogel, 2010; Farmer & Vogel, 2013). The scaling has been conducted via a simple least square linear regression (e.g., Sachindra et al., 2013) as follows,

$$\widehat{\mathbf{Q}}_\mathbf{j} = \gamma_{0j} + \gamma_{ij} \cdot \mathbf{Q_i} \tag{1}$$

where $\boldsymbol{\gamma_{ij}}$ and $\boldsymbol{\gamma_{0j}}$ are the regression coefficients of the slope and intercept, respectively.

### 2.1.2 Multiple donor gauges

When multiple donor gauges are used, a multiple linear regression (MLR) is a natural approach for estimating the streamflow at the target gauge. Here, the discussion goes beyond the inference application using multi-donor gauges and is extended to lay out the foundation for the Graphical Lasso (Glasso) algorithm of this study. A generalization is thus made herein that given a set of total $p$ gauges with daily streamflow records, any one of them may serve as a potential target gauge while the rest of p-1 gauges as the donors. Estimation of the streamflow at the selected target, $j$, by MLR using all available p-1 donor gauges can be expressed as follows,

$$\widehat{\mathbf{Q}}_\mathbf{j} = \eta_{0j} + \sum_{i=1, i \neq j}^{p} \eta_{ij} \cdot \mathbf{Q_i} \tag{2}$$

where $\eta_{ij}$ and $\eta_{0j}$ represent the multiple regression coefficients.

Since the probability distribution of streamflow, $\boldsymbol{Q}$, is often well approximated by a log-normal distribution (e.g., Stedinger, 1980), equation (2) can be modified and written as,

$$\widehat{\mathbf{Y}}_\mathbf{j} = \rho_{0j} + \sum_{i=1, i \neq j}^{p} \rho_{ij} \cdot \mathbf{Y_i} \tag{3}$$

where $\mathbf{Y}$ follows a normal distribution with $\mathbf{Y} = \ln(\boldsymbol{Q})$, and $\rho_{ij}$ and $\rho_{0j}$ represent the multiple regression coefficients. To avoid numerical issues with the logarithm of zero-valued streamflow, Farmer (2016) assigned a small constant value (e.g. 0.00003 m³/s) to replace zero. Here we add 1 to it instead so that $\mathbf{Q} = 0$ comes out as $\mathbf{Y} = 0$ as well,

$$\mathbf{Y_i} = ln(\mathbf{Q_i} + 1) \tag{4}$$

The results of applying either Farmer's approach or equation (4) are almost identical as shown in this study (see Section 3.1.2).

For convenience and simplicity, $\mathbf{Z}$, the standard Gaussian form of $\mathbf{Y}$ is instead used, i.e.,

$$\mathbf{Z_i} = \frac{\mathbf{Y_i} - \mu_{y_i}}{\sigma_{y_i}} \tag{5}$$

The corresponding regression equation thus has zero intercept:

$$\hat{\mathbf{Z}}_{\mathbf{j}} = \sum_{i=1,i \neq j}^{p} \mathbf{Z_i} \cdot \alpha_{ij} \tag{6}$$

It can be further written in a matrix form by posing $\mathbf{Z}$ as the join of the column vectors, $(\mathbf{z_1}, \ldots, \mathbf{z_p})$, as follows,

$$\hat{\mathbf{Z}} = \mathbf{Z} \cdot \mathbf{A} \tag{7}$$

where $\mathbf{A}$ is the $p$ by $p$ regression coefficient matrix,

$$\mathbf{A} = \begin{bmatrix} 0 & \alpha_{12} & \cdots & \alpha_{1p} \\ \alpha_{21} & 0 & \cdots & \alpha_{2p} \\ \cdots & \cdots & 0 & \cdots \\ \alpha_{p1} & \cdots & \cdots & 0 \end{bmatrix} \qquad (8)$$

$\hat{\mathbf{Z}}$ and $\mathbf{Z}$ are $n$ by $p$ matrices where $n$, as defined before, is the number of daily streamflow records. Note that if the streamflow data do not follow the log-normal distribution as assumed, one can easily transform the data into the log-normal distribution so that equations (3)-(8) are applicable.

## 2.2 Covariance and precision matrices

Two methods used in computing matrix $\mathbf{A}$ are described herein. The first method is based on the covariance matrix $\mathbf{\Sigma}$ and the second one, which forms the core step of Graphic Lasso, uses the inverse of the covariance matrix, which is called the precision matrix, $\mathbf{\Theta}$, i.e.,

$$\mathbf{\Theta} = \mathbf{\Sigma}^{-1} \qquad (9)$$

Since the true covariance ($\mathbf{\Sigma}$) or precision ($\mathbf{\Theta}$) matrices are unknown, estimates are used, in which an estimated covariance matrix is denoted by $\mathbf{W}$, and from it an estimated precision matrix is denoted by $\hat{\mathbf{\Theta}}$. Following Friedman et al. (2008), the columns and rows of $\mathbf{W}$ can be permuted so that the target gauge $j$ is the last and thus it facilitates a partition of the $\mathbf{W}$ matrix into four blocks consisting of a square submatrix $\mathbf{W_{11}}$ of size $(p-1) \times (p-1)$, a column vector $\mathbf{w_{12}}$ of $1 \times (p-1)$, its transpose $\mathbf{w_{12}^T}$, and a scalar $w_{22}$. The same partition is also applied to $\hat{\mathbf{\Theta}}$ giving the four blocks of $\hat{\mathbf{\Theta}_{11}}, \hat{\mathbf{\Theta}_{12}}, \hat{\mathbf{\Theta}_{12}^T}$ and $\hat{\theta}_{22}$, respectively. The identity matrix $\mathbf{I}$, $\mathbf{W} \cdot \hat{\mathbf{\Theta}} = \mathbf{I}$, can then be written as (Friedman et al., 2008),

$$\begin{pmatrix} \mathbf{W_{11}} & \mathbf{w_{12}} \\ \mathbf{w_{12}^T} & w_{22} \end{pmatrix} \begin{pmatrix} \hat{\mathbf{\Theta}_{11}} & \hat{\mathbf{\Theta}_{12}} \\ \hat{\mathbf{\Theta}_{12}^T} & \hat{\theta}_{22} \end{pmatrix} = \begin{pmatrix} \mathbf{I} & \mathbf{0} \\ \mathbf{0}^T & 1 \end{pmatrix} \qquad (10)$$

By decomposing $\mathbf{A}$ into column vectors as follows,

$$\mathbf{A} = [\boldsymbol{\alpha_1} \quad \cdots \quad \boldsymbol{\alpha_j} \quad \cdots \quad \boldsymbol{\alpha_p}] \tag{11}$$

a solution based on the covariance matrix can be obtained by,

$$\boldsymbol{\alpha_j} = \mathbf{W_{11}^{-1}} \cdot \mathbf{w_{12}} \tag{12}$$

In contrast, the solution using the precision matrix which does not involve matrix inverse but a simple division can be obtained by,

$$\boldsymbol{\alpha_j} = -\frac{1}{\hat{\theta}_{22}}\widehat{\boldsymbol{\theta}}_{\mathbf{12}} \tag{13}$$

Equation (13) is derived by making use of $\mathbf{W_{11}} \cdot \widehat{\boldsymbol{\theta}}_{\mathbf{12}} + \mathbf{w_{12}} \cdot \hat{\theta}_{22} = \mathbf{0}$ from equation (10) and equation (12). Equation (13) shows how the precision matrix $\widehat{\boldsymbol{\theta}}$ and the matrix of regression coefficients, $\mathbf{A}$, are related to each other. The elements of the matrix $\mathbf{A}$ are computed as $\boldsymbol{\alpha_j}$ for $1 \leq j \leq p$. The vector of regression coefficients $\boldsymbol{\alpha_j}$ for the $j$th column of $\mathbf{A}$ is proportional to the vector $\widehat{\boldsymbol{\theta}}_{\mathbf{12}}$ of $\widehat{\boldsymbol{\theta}}$.

If in the place of $\mathbf{W}$, the sample covariance matrix $\mathbf{S}$ is used, with its corresponding precision matrix estimate, $\mathbf{T}$, the coefficient $\boldsymbol{\alpha_j}$ in matrix $\mathbf{A}$ can then be expressed by,

$$\boldsymbol{\alpha_j} = \mathbf{S_{11}}^{-1} \cdot \mathbf{s_{12}} \tag{14}$$

$$\boldsymbol{\alpha_j} = -\frac{1}{t_{22}}\mathbf{t_{12}} \tag{15}$$

where $\mathbf{S_{11}}, \mathbf{s_{12}}, \mathbf{t_{12}}$, and $t_{22}$ correspond to $\mathbf{W_{11}}, \boldsymbol{w_{12}}, \widehat{\boldsymbol{\theta}}_{\mathbf{12}}$, and $\hat{\theta}_{22}$, respectively, and $\mathbf{S}$ and $\mathbf{T}$ are,

$$\mathbf{S} = \frac{1}{n-1}\mathbf{Z^T} \cdot \mathbf{Z} \tag{16}$$

$$\mathbf{T} = \mathbf{S^{-1}} \tag{17}$$

## 2.3 Concept and construction of Gaussian graphical models

A graph $\mathbf{G}$ ($\mathbf{v}$, $\mathbf{e}$) is used to describe the correlation topology of a streamflow gauge network in our graphical Markov modeling. Let $\mathbf{G}$ ($\mathbf{v}$, $\mathbf{e}$), represented by a set of vertices $\mathbf{v}$ (i.e., gauges in our model) and a set of edges $\mathbf{e}$ with each edge connects two vertices, define a conditional independence structure of a given streamflow gauge network. Namely, if there is no edge between any two vertices, then these two vertices are conditionally independent from each other given the remaining connections in the graph. Thus, a graph $\mathbf{G}$ ($\mathbf{v}$, $\mathbf{e}$) so constructed represents a global view of the dependence structure of all the vertices in the graph. That is, if there is an edge between two gauges, the streamflows between these two gauges are conditionally dependent on each other, otherwise conditionally independent given the remaining gauges' dependency in the network. It follows that a gauge can only be a donor gauge if it is connected to the target gauge with an edge. Because the streamflow data are converted into standard Gaussian vectors, $\mathbf{Z}$, a graph model for the streamflow network, $\mathbf{G}$, built on top of these gauges, is a *Gaussian graphical model*. If each gauge in the network depends (conditionally) on all of the remaining gauges in the hydrometric network, then, all of the gauges are connected. That is, each vertex is connected to all of the remaining vertices in the graphical modeling network. Such a network graph is called a full or complete graph with $\frac{p^2-p}{2}$ edges. From the graphical modeling concept point of view, the key to find the donor gauges for each target gauge in the streamflow network is to first construct the network graph $\mathbf{G}$ in such a way that $\mathbf{G}$ defines a conditional independence structure. Our insight is that any graph constructed based on the pair-wise correlations does not define a conditional independence structure. Therefore, just considering the pair-wise correlation would result in a misleading situation that more donor gauges than necessary appear to be needed for each selected target gauge in the network. This is why in the past when pair-wise correlations were used, often a high threshold value has to be used to sort of walk around this issue, or simply a fixed number of donor gauges are decided *a priori*, such as, 1, 2, or 3 donor gauges, for each target gauge. It should be emphasized here that the pair-wise correlations are computed directly from the covariance matrix, which does not reflect the important global conditional independence structure among the gauges in the network. Therefore, using the covariance matrix cannot provide an effective pathway to the donor gauge selection. In contrast, the precision matrix, an inversion of the covariance matrix, reveals the global conditional independence structure among the gauges in the network (Friedman et al., 2008; Koller & Friedman, 2009). If discharges at two gauges $i$ and $j$ are conditionally independent, then $\hat{\theta}_{ij} = 0$ in the estimated precision matrix, $\hat{\mathbf{\Theta}}$, and otherwise $\hat{\theta}_{ij} \neq 0$. Indeed, the gauge network graph, $\mathbf{G}$, should be contructed based on the precision matrix, $\hat{\mathbf{\Theta}}$, instead of the covariance matrix $\mathbf{W}$. A graph $\mathbf{G}$ can be represented by an adjacency matrix defined as follows,

$$\mathbf{G} = \begin{cases} g_{ij} = 1 & if \quad |\hat{\theta}_{ij}| > 0 \\ g_{ij} = 0 & otherwise \end{cases} \tag{18}$$

where $g_{ij}$ and $\widehat{\theta}_{ij}$ represent the element of the *ith* row and *jth* column of $\mathbf{G}$ and $\hat{\mathbf{\Theta}}$, respectively, $g_{ij} = 1$ means the gauges $i$ and $j$ are linked or adjacent. In practice, both the estimated covariance matrix $\mathbf{W}$ and the estimated precision matrix $\hat{\mathbf{\Theta}}$ are approximated by the sample covariance matrix $\mathbf{S}$ defined by equation (16) and the sample precision matrix $\mathbf{T}$, defined by equation (17). The graph $\mathbf{G}$ thus obtained with the sample precision matrix $\mathbf{T}$ generally is dense (d'Aspremont et al., 2008) due to the nature of the noisy data. An unnecessary dense graph, especially the one caused by noisy data, is not useful since it provides no distinction in relevance.

As the number of gauges, $p$, in graph **G** increases, the complexity of the graph grows rapidly as the number of edges grows rapidly. The graphical Lasso is used in trimming a graph sparse and makes it useful.

## 2.4 The Graphical Lasso

The *Graphical Lasso* (Glasso) is an algorithm developed initially by Friedman et al. (2008) which imposes sparsity on the precision matrix by tuning a regularization parameter $\lambda$. This algorithm has been actively used, analyzed and improved by several authors (Mazumder & Hastie, 2012; Sojoudi, 2014; Witten et al., 2011). Our work used the Glasso Matlab package (*glasso*) and also a more recent efficient implementation called *GLASSOFAST* (Sustik & Calderhead, 2012).

The *Glasso* algorithm provides an efficient solution by maximizing the Gaussian log-likelihood according to the formulation given in equation (19), adapted from Friedman et al. (2008), where *det* and *tr* are the determinant and trace of a square matrix respectively, $||\mathbf{\Theta}||_1$, is the $L_1$ norm of the precision matrix $\mathbf{\Theta}$ (i.e., the sum of the absolute value of all the elements in the matrix) and $\lambda$ is the $L_1$ norm regularization parameter.

$$\widehat{\mathbf{\Theta}}_{\text{Glasso}} \equiv arg_{\mathbf{\Theta}} max[log(det\mathbf{\Theta}) - tr(\mathbf{S} \cdot \mathbf{\Theta}) - \lambda||\mathbf{\Theta}||_1] \tag{19}$$

The *Glasso* algorithm requires that the probability distribution of the input data be relatively well described by a multivariate Gaussian distribution as in our case for **Z**. The inputs required by the *Glasso* algorithm are the empirical covariance matrix **S** and a regularization parameter $\lambda$. The output from the *Glasso* algorithm is a potentially sparse precision matrix estimate, $\widehat{\mathbf{\Theta}}_{\text{Glasso}}$. Equation (20) shows the inputs and output of the *Glasso* algorithm expressed as,

$$\widehat{\mathbf{\Theta}}_{\text{Glasso}} = Glasso(\mathbf{S}, \lambda) \tag{20}$$

For estimating the target streamflow time series using MLR after donor gauge selection, the regression coefficients of matrix $\mathbf{A}$ can be found from $\widehat{\mathbf{\Theta}}_{\text{Glasso}}$ by applying $\widehat{\mathbf{\Theta}}_{\text{Glasso}}$ to equation (13) in which $\widehat{\mathbf{\Theta}}$ is replaced by $\widehat{\mathbf{\Theta}}_{\text{Glasso}}$ as shown in equation (21) below,

$$\mathbf{\alpha_j} = -\frac{1}{\widehat{\theta}_{Glasso_{22}}}\widehat{\mathbf{\Theta}}_{Glasso_{12}} \tag{21}$$

If the regularization parameter $\lambda$ is equal to zero, no regularization is imposed, and the estimated precision matrix $\widehat{\mathbf{\Theta}}_{\text{Glasso}}$ is equivalent to $\mathbf{T} = \mathbf{S^{-1}}$ and the corresponding graph $\mathbf{G}$ remains not sparse. On the other hand, if the regularization parameter is a large value, $\widehat{\mathbf{\Theta}}_{\text{Glasso}}$ will be over-regularized, and the underlying graph $\mathbf{G}$ would have no edges. Given the importance of the $\lambda$ parameter, we devise a new algorithm called *Selection of Graph Model* (SGM) in this study to obtain an optimal graph, which is presented in Section 3.2. SGM is to select the $\lambda$ parameter based on a multi-objective optimization procedure that minimizes the number of edges of the underlying Gaussian graphical model and also the errors in streamflow estimates. The objective is to keep a good balance between the complexity of a graph and the accuracy of an estimation for the target gauge. This is achieved by promoting sparsity while minimizing the estimation error.

## 3. New Approach of Selecting Multiple Donor Gauges via Graphical Models

Given a target gauge, our new approach objectively selects a set of necessary gauges from all the gauges in a streamflow network as donor gauges via the sparse graph model built for the network. Our new donor selection algorithm comprises two intertwined procedures: First, we apply the Glasso method on the precision matrix to trim the graph sparse. Second, we adopt an estimator, such as a multiple linear regression method, to obtain the streamflow estimation errors. The final sparse graph is determined by balancing between the graph sparsity and the estimation errors. Such a balance is achieved by minimizing two main objectives: (1) the model complexity and (2) the inferred streamflow estimation error. Here, model complexity refers to the number of edges included in the graph. Thus, the simplest model would be a graph with the smallest number of edges with which there is only one single donor gauge for each target gauge, while the most complex model would be a complete graph with which all of the available gauges in the network become the donor gauges for each target gauge. We argue, and show in the example applications, that there is an appropriate trade-off between the model complexity and the accuracy of the inferred streamflow estimates, and that a more complex model is not necessarily better than a simpler model due to noises and relevance. The balance between sparsity and estimation error is achieved through a multiple objective optimization using the Pareto front. To minimize the streamflow estimation error, we use a regression method for the flow estimates. We use this inference method herein because it is simple and computational efficient. In practice, one can certainly apply any other available inference methods that one prefers. At the end of the iteration, the gauges that share their edges with the target gauge in the final graph are identified as the donor gauges.

In summary, the essence of our new method is that it explicitly and effectively considers the correlation structure of the entire gauge network, based on conditional independence structure, embodied by the underlying streamflow network graph **G** in our graphical modeling, rather than the pair-wise correlation between any two gauges as used in the existing methods. When the final sparse graph is achieved, the gauges that have edges to the target gauge are identified as the donor gauges for the given target gauge. As the donor gauges are selected based on the conditional independence structure of the entire gauge network graph, **G**, our method is a global approach as opposed to the existing pair-wise methods which can be viewed as a local approach. Also, if there are multiple target gauges for the study streamflow network, our method provides the donor gauges for each target gauge in the network simultaneously. One does not need to select the donor gauges one at a time for each target gauge. Major steps involved in the donor gauge selection process via the graph model are summarized in Figure 1.

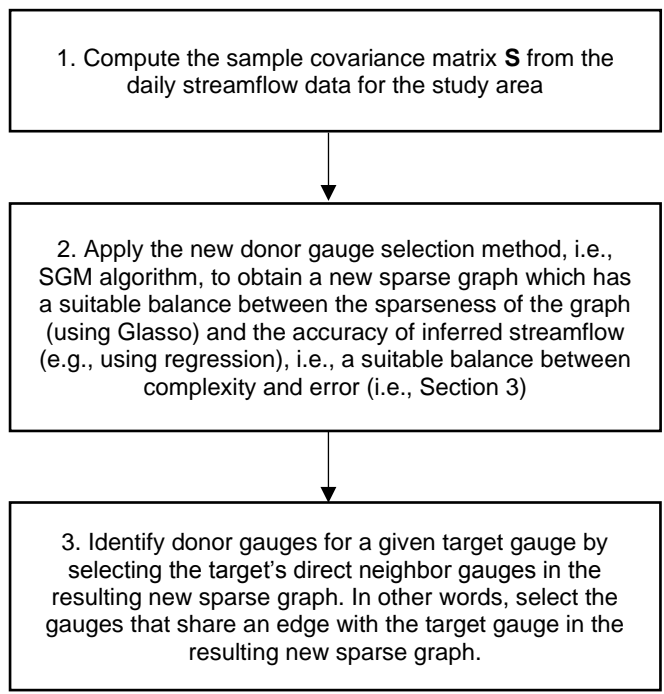

Figure 1. A diagram for the new donor gauge selection method with major steps in the overall process of building the graph and pruning it.

### 3.1 Donor gauge selection via Graphical Model

Our approach in selecting a proper set of donor gauges to be used for inferring streamflow for each target gauge is to exploit the conditional independence structure encoded in the precision matrix. This is accomplished by promoting sparsity on the precision matrix that results in fewer edges of the underlying graph **G**. This is consistent with the parsimonious principle. A simpler model that explains well the observations should be preferred over more complex models. Under such a context, the parsimonious principle implies a selection of an underlying graphical model that is as sparse as possible while keeping the estimation error relatively low.

### 3.1.1 Imposition of sparsity to underlying graphical model

The sparsity is achieved by adjusting the regularization parameter $\lambda$ for the *Glasso* algorithm in conjunction with a thresholding procedure that uses a truncation parameter $\tau$, which modifies equation (18) as follows,

$$\mathbf{G}_\tau = \begin{cases} g_{ij} = 1 & if \quad |\hat{\theta}_{ij}| > \tau \\ g_{ij} = 0 & otherwise \end{cases} \tag{22}$$

The thresholding procedure is required in addition to the $L_1$ norm regularization because even though the $L_1$ norm of the precision matrix decreases monotonically as $\lambda$ increases, the number of edges in the graph $G$ does not necessarily decrease monotonically. Therefore, a multi-objective optimization is needed to simultaneously minimize the mean error between the observed random variable $Z$ and the inferred data matrix $\hat{Z}$ from equation (7), and the number of edges of the underlying graph $G$. There are some situations where a particular edge has to be removed from the underlying graphical model by setting the element $g_{ij}$ to zero. One example of such situation is when both gauges are known to be donor gauges, therefore none of them will be inferred from each other and the corresponding edge in the graphical model should be removed. A similar case arises when both gauges are known to be the target gauges, the edge between them does not help in this situation, as one gauge should not be infered using the other as a donor. In implementing the Glasso procedure, representing by Equation (23), an optional input parameter, graph $G_\tau$, facilitates removal of some edges. If that graph $G_\tau$ is omitted in the input, all edges are available. The implementation also allows one to compute the sparse precision matrix with a prescribed sparsity pattern. Through the process of pruning the original dense graph, the donor gauges are identified as those which are directly linked to the target gauge in the resulting sparse graph.

$$\widehat{\Theta}_{train(\lambda, G_\tau)} = Glasso\left(\mathbf{S}_{train}, \lambda, \boldsymbol{G}_{\tau,}\right) \tag{23}$$

### 3.1.2 Estimation of streamflow errors

The normalized standard Gaussian daily streamflow data set, $\mathbf{Z}$, is sorted chronologically and then divided into three disjoint subsets of approximately same size. The subsets are used for training, validation and testing and designate, respectively, as $\mathbf{Z}_{train}$, $\mathbf{Z}_{val}$, and $\mathbf{Z}_{test}$. $\mathbf{Z}_{train}$ is used for training the inference model by computing the regression coefficients in matrix $A$. $\mathbf{Z}_{val}$ is used for choosing the $\lambda$ and $\tau$ values that minimize the validation error and the number of edges of the underlying graph $\boldsymbol{G}$. $\mathbf{Z}_{test}$ is used for assessing the predictive capability of the streamflow using the new donor gauge selection method together with the MLR inference method through estimating the error based on the $\mathbf{Z}_{test}$ data. It is worth mentioning that accuracy of the inferred daily streamflow estimates depends on how well donor gauges are selected; and that an inference approach can be selected with much freedom. The first two thirds of the daily streamflow records are randomly assigned to the training $\mathbf{Z}_{train}$ and validation $\mathbf{Z}_{val}$ data sets with a split ratio of 50%. The remaining one third of the most recent data are used as the test set $\mathbf{Z}_{test}$. Streamflow errors are estimated based on the following process:

### (1) Estimation of training covariance and sparse precision matrices

The training precision matrix, $\widehat{\Theta}_{train(\lambda, G)}$, under a given value of the regularization parameter $\lambda$, is computed in the *Glasso* algorithm of equation (20) using its sample variance $\mathbf{S}_{train}$. The initial sparsity of the training precision matrix, $\widehat{\Theta}_{train(\lambda, G)}$, is determined by the regularization parameter $\lambda$. Additional sparsity is achieved for a given value of the truncation parameter $\tau$ defined in equation (22). A new training precision matrix $\widehat{\mathbf{\Theta}}_{train(\lambda,G_\tau)}$ is then computed using equation (23) and the sparse graph $\mathbf{G}_\tau$. This sparse precision matrix has a value of zero on all elements where the graph $\mathbf{G}_\tau$ has missing edges.

(2) Estimation of regression coefficients and streamflow

$\widehat{\mathbf{\Theta}}_{train(\lambda,G_\tau)}$ is used in obtaining the regression coefficient matrix, $\mathbf{A}_{train}$, with equation (21). The standardized validation streamflow time series, $\widehat{\mathbf{Z}}_{val}$, are then estimated using $\mathbf{A}_{train}$ and the validation dataset, $\mathbf{Z}_{val}$, as below,

$$\widehat{\mathbf{Z}}_{val} = \mathbf{Z}_{val} \cdot \mathbf{A}_{train} \tag{24}$$

The estimated log-transformed validation streamflow data, $\widehat{\mathbf{Y}}_{val}$, is calculated from $\widehat{\mathbf{Z}}_{val}$. That is, $\widehat{\mathbf{Y}}_{val_j}$ is computed as follows,

$$\widehat{\mathbf{Y}}_{val_j} = \widehat{\mathbf{Z}}_{val_j} \cdot \sigma_{y_{val_j}} + \mu_{y_{val_j}} \tag{25}$$

where $\mu_{y_{val_j}}$ and $\sigma_{y_{val_j}}$ are, respectively, the mean and standard deviation of $\widehat{\mathbf{Y}}_{val_j}$. The estimated validation streamflow data, $\widehat{\mathbf{Q}}_{val}$, is then obtained by,

$$\widehat{\mathbf{Q}}_{val_j} = exp\left(\widehat{\mathbf{Y}}_{val_j}\right) - 1 \tag{26}$$

(3) Score function and validation error

Selection of the graphical model should maximize the quality of the inferred daily streamflow time series, that is, the daily streamflow time series estimates at the target site should be as accurate as possible. A score function is designed to measure the accuracy of the inferred values. Equation

(27) defines a conditional goodness-of-fit metric that calculates the value of the coefficient of determination, $R^2$, between the observed and estimated daily streamflow time series for the validation data set, where $R^2_{val_j}$ is the $R^2$ between the observed streamflow $\mathbf{Q}_{val_j}$ used for validation and the estimated streamflow $\widehat{\mathbf{Q}}_{val_j}$, for $1 \leq j \leq q$, where $j$ is an index representing the $jth$ gauge and $q$ is the number of target gauges. By default, all of the gauges are considered as potential target sites, where $q$ is equal to $p$. The score is positive if $\boldsymbol{R}^2_{val_j}$ is greater than an assigned threshold $\varGamma$, otherwise, it is taken as zero. Equation (28) calculates the validation score. Equation (29) defines the validation error, ranging within [0, 1], that is used in our multi-objective optimization procedure.

$$score_{val_j} = \begin{cases} R^2_{val_j} = R^2\left(\mathbf{Q}_{val_j}, \widehat{\mathbf{Q}}_{val_j}\right) & if \quad R^2_{val_j} > \varGamma \\ 0 & otherwise \end{cases} \tag{27}$$

$$score_{val} = \sum_{j=1}^{q} score_{val_j}, \quad q \leq p \tag{28}$$

$$error_{val} = \frac{q - score_{val}}{q} \tag{29}$$

The validation error of the graphical model is selected in such a way that it will maximize the validation score thus minimize the error.

## 3.2 Donor selection: Selection of Graph Model algorithm

On this basis of the proceeding procedures, we devise an algorithm called *Selection of Graph Model* (SGM) to obtain an optimal underlying graph. A graph determined by the SGM algorithm is denoted as $\boldsymbol{G}_{sgm}$. The SGM algorithm implements a multi-objective optimization procedure where the optimization objectives include: (1) minimizing the number of edges of the underlying graph to make it as sparse as possible, and (2) minimizing validation error. SGM generates a set of regularization parameter $\lambda$ within the range of $[\lambda_{min}, \lambda_{max}]$. For each $\lambda$, a truncation parameter, $\tau$, is selected such that the underlying graph has the given number of edges between $[K_{min}, K_{max}]$. Given the multi-objective nature of the problem, a set of graphs corresponding to a set of non-dominated solutions on the Pareto front are obtained as a result. A graph $\boldsymbol{G}_{sgm}$ thus represents one of the graphs from the set. A final graph or a graph set, $\{\boldsymbol{G}_{sgm}\}$, is selected from the set of candidate solutions by balancing the trade-offs between the model complexity and error.

The pseudo code for SGM algorithm is provided in the Algorithm 1. The parameter *res* is an integer that represents the resolution of a sequence of sampling values to create *res* number of **lamba_set** vector with values between $\lambda_{min}$ and $\lambda_{max}$. *DonorSet* and *TargetSet* are optional parameters that represent a set of identifiers of the gauges that are known to be donors or targets, respectively. The defaults for *DonorSet* and *TargetSet, in Algorithm 1,* are empty sets. That is, any gauge can potentially be used as a *Donor* or *Target* gauge. If *DonorSet* or *TargetSet* are non-null sets, then their corresponding gauges are treated as donor or target gauges, respectively. For them, the graph model $\boldsymbol{G}_{\tau}$ defined in equation (22) removes all the edges between the *ith* and *jth* gauge when both, *i* and *j,* belong to *DonorSet* or both belong to *TargetSet.* The *getSequence* function generates the **lamba_set** vector, which can be generated using a linear sequence.

**Algorithm 1:** *Selection of Graph Model (SGM)*

---

**STEP 0.** Define the SGM inputs (assignment of default values)

$\lambda_{min} = 0.01;\; \lambda_{max} = 0.10;\, K_{min} = 10;\, K_{max} = \frac{p^2 - p}{2};\;\; res = 30;\, \Gamma = 0.7$

*DonorGroup:={}; TargetGroup:={}*

Retrieve training ($\mathbf{Z}_{train}$) and validation ($\mathbf{Z}_{val}$) data sets;

**STEP 1**. Compute the empirical covariance matrix using equation (16) from the training set:

$n_{train} = length(\mathbf{S}_{train})$

$\mathbf{S}_{train} = \frac{1}{n_{train}-1}\mathbf{Z}_{train}{}^{\mathsf{T}} \cdot \mathbf{Z}_{train};$

**STEP 2.** Generate Multi-objective optimization sampling points:

> ***lambda_set*** = *getSequence(minVal=$\lambda_{min}$, maxVal = $\lambda_{max}$, res);*
>
> ***for** r=1 **to** res:*
>
>> $\lambda_r$= ***lambda_set[r];***
>>
>> Compute the initial precision matrix from $\mathbf{S}_{train}$ using equation (20):
>>
>> $\widehat{\boldsymbol{\Theta}}_{train_r} = Glasso(\mathbf{S}_{train}, \lambda_r);$
>>
>> ***for** k= $K_{min}$ **to** $K_{max}$:*
>>
>>> choose $\tau_{r,k}$ to compute the underlying graph model with at most $k$ edges using equation (22):
>>>
>>> $\mathbf{G}_{r,k} = \begin{cases} g_{r,k_{ij}} = 1 & if \quad |\widehat{\theta}_{train_{r_{ij}}}| > \tau_{r,k} \\ g_{r,k_{ij}} = 0 & otherwise \end{cases};$
>>>
>>> Compute the sparse training precision matrix, using equation (23):
>>>
>>> $\widehat{\boldsymbol{\Theta}}_{train_{r,k}} = Glasso(\mathbf{S_{train}}, \lambda_r, \mathbf{G}_{r,k});$
>>>
>>> Compute the training matrix of regression coefficients $\mathbf{A}_{train_{r,k}}$ from $\widehat{\boldsymbol{\Theta}}_{train_{r,k}}$, using equation (21), for $1 \le j \le p$:
>>>
>>> $\boldsymbol{\alpha}_{train_{r,k_j}} = -\frac{1}{\widehat{\theta}_{train_{r,k_{22}}}}\widehat{\boldsymbol{\theta}}_{train_{r,k_{12}}};$
>>>
>>> Compute the inferred Z-score log-transformed validation streamflow from equation (24):
>>>
>>> $\widehat{\mathbf{Z}}_{val_{r,k}} = \mathbf{Z}_{val} \cdot \mathbf{A}_{train_{r,k}};$
>>>
>>> Compute the inferred log-transformed validation streamflow using equation (25) for $1 \le j \le p$:
>>>
>>> $\widehat{\mathbf{Y}}_{val_{r,k_j}} = \widehat{\mathbf{Z}}_{val_{r,k_j}} \cdot \sigma_{y_{val_j}} + \mu_{y_{val_j}};$
>>>
>>> Compute the inferred validation streamflow using equation (26) for $1 \le j \le p$:
>>>
>>> $\widehat{\mathbf{Q}}_{val_{r,k_j}} = exp\left(\widehat{\mathbf{Y}}_{val_{r,k_j}}\right) - 1;$
>>>
>>> Calculate the validation score using equation (27) and equation (28) for $1 \le j \le q$:
>>>
>>> $score_{val_{r,k_j}} = \begin{cases} \mathrm{R}^2_{val_j} = \mathrm{R}^2\left(\mathbf{Q}_{val_j}, \widehat{\mathbf{Q}}_{val_j}\right) & if \quad \mathrm{R}^2_{val_j} > \Gamma \\ 0 & otherwise \end{cases}$
>>>
>>> $score_{val_{r,k}} = \sum_{j=1}^{q} score_{val_{r,k_j}}, \quad q \le p;$
>>>
>>> Calculate the validation error using equation (29):
>>>
>>> $error_{val_{r,k}} = \frac{q - score_{val_{r,k}}}{q};$
>>>
>>> store the sampling results: *multi_objective_points* = [$k$, $error_{val_{r,k}}$], $\lambda_r$ and $\mathbf{G}_{r,k}$.

**STEP 3.** Select the set of non-dominated solutions from *multi_objective_points*

**STEP 4.** From the set of non-dominated solutions, select a sparse graph (as the output), $\boldsymbol{G}_{sgm}$, with a suitable tradeoff between the number of edges and validation error and optionally the corresponding matrix of regression coefficients $\boldsymbol{A}_{sgm}$.

## 4. Application Examples

To illustrate the proposed method, two example applications are presented. One is about the stream flow inference; and the other the removal of streamflow gauges with the least loss of information. The problem involved and the solution procedures are given herein, but the study area, the data and problem setup are given in the next section. This section starts at the completion of the Algorithm 1 where the optimized underlying graph or graph set $\{G_{sgm}\}$ has already been obtained.

### 4.1 Donor gauge selection and flow inference example

The donor gauge selection task is greatly simplified once the underlying graph $G_{sgm}$ is identified by the SGM algorithm as this graph $G_{sgm}$ reveals conditional independent conditions between the streamflow gauges for a given hydrometric streamflow network. A set of donor gauges best for each target gauge is explicitly contained in the graph $G_{sgm}$. That is, a best set of donor gauges for a given target gauge is the set of gauges that are directly connected to it in the graph, $G_{sgm}$. Such a set of donor gauges includes only those on which each target station depends. We note that once the donor gauges are identified based on our new donor gauge selection algorithm (i.e., SGM), one can apply any inference method, including the advanced machine learning methods, to estimate the streamflow for the target gauges. In this study, we apply a regression-based inference method described in Section 2 due to its simplicity, for the objective of its use here is to form the basis for comparisons. In all comparisons with the other different donor gauge selection approaches in this study, the inference method used is identical.

### 4.1.1 Inference of daily streamflow time series with graph $\boldsymbol{G}_{sgm}$

In this application example, the streamflow data at gauge sites are divided into three sets as previously described. We further consider that each gauge in the network could become a target gauge with its latest one third of data missing. The inference of daily streamflow is thus performed for every single gauge in the study area.

With the selected donor gauges taken from graph $\boldsymbol{G}_{sgm}$, there are three ways in obtaining the streamflow time series $\widehat{\mathbf{Q}}_{test}$ based on the MLR method. Each is slightly different from one another in how the regression coefficients are obtained. The first approach directly applies the MLR to the normalized Gaussian variable $\mathbf{Z}_{test}$ as follows. Let matrix $\boldsymbol{A}_{sgm}$ represent matrix $\mathbf{A}$ of equation (8) whose element $\alpha_{ij}$ is determined based on graph $\boldsymbol{G}_{sgm}$. The log-transformed streamflow time series for the test set $\widehat{\mathbf{Z}}_{test}$ can then be estimated directly using matrix $\boldsymbol{A}_{sgm}$ and the test dataset $\mathbf{Z}_{test}$ as follows,

$$\widehat{\mathbf{Z}}_{test} = \mathbf{Z}_{test} \cdot \boldsymbol{A}_{sgm} \tag{30}$$

To obtain $\widehat{\mathbf{Y}}_{test}$ from $\widehat{\mathbf{Z}}_{test}$, the unknown mean $\mu_{Y_{test_j}}$ and standard deviation $\sigma_{Y_{test_j}}$ for the test set are assumed to be the same as those for the training set. Then, the streamflow time series $\widehat{\mathbf{Q}}_{test}$ can be obtained from $\widehat{\mathbf{Y}}_{test}$. The weakness of this approach is that the assumption made here is not usually held.

The second approach applies the MLR over the log-transformed streamflow time series for the training data set over $1 \leq j \leq p$, using only the donors for the *jth* target site as expressed by equation (31),

$$\widehat{\mathbf{Y}}_{\text{test}_j} = \beta_{0j} + \sum_{i=1}^{size(donors(j))} \beta_{ij} \cdot \mathbf{Y}_{train_{donors(j)_i}} \tag{31}$$

where $donors(j) = donors(\boldsymbol{G}_{sgm}, j)$, $\beta_{ij}$ and $\beta_{0j}$ are the regression coefficients. The daily streamflow time series, $\widehat{\mathbf{Q}}_{\text{test}_j}$, is then estimated from $\widehat{\mathbf{Y}}_{\text{test}_j}$ as follows,

$$\widehat{\mathbf{Q}}_{\text{test}_j} = exp\left(\widehat{\mathbf{Y}}_{\text{test}_j}\right) - 1 \tag{32}$$

The third approach simply applies the MLR to the non-transformed streamflow time series avoiding the logarithmic transformation. Among these three approaches, we used the second one with equations (31-32). As pointed out by Farmer (2016), this approach generally produced either more accurate or more stable results than the third approach.

### 4.1.2 Building base graphs for distance- and correlation-based approaches for comparisons

To evaluate the performance of our new donor gauge selection method based on graph $\boldsymbol{G}_{sgm}$ in inferring daily streamflow time series, we compare our new method with two widely used donor gauge selection methods, the distance-based method ("Dist") and the pair-wise correlation-based method ("Corr"). These two selection methods have also been shown in the literature to be effective and are evaluated here for comparison purpose. To put into a consistent framework for fair comparison, we construct two graphs: $\boldsymbol{G}_{dist}$ (distance-based) and $\boldsymbol{G}_{corr}$ (pair-wise correlation-based). Starting from an empty graph with no edges, the $\boldsymbol{G}_{dist}$ graph is built by adding to each target site a link to its nearest neighbor site. In this case, each target site has one donor site, expressed as $\boldsymbol{G}_{dist,1}$, and the constructed graph structure is determined by the number of edges added and their relative locations in the gauge network. For the case of each target having $m$ donor gauges, edges between each target site and its nearest *m* neighbor gauges are added in the graph. Here we consider only up to three donor gauges for a target gauge, that is, graph $\boldsymbol{G}_{dist,2}$ and $\boldsymbol{G}_{dist,3}$. The graph of $\boldsymbol{G}_{corr}$ is built in a similar way to $\boldsymbol{G}_{dist}$ except that the highest pair-wise correlation is used, forming $\boldsymbol{G}_{corr,1}$, $\boldsymbol{G}_{corr,2}$ and $\boldsymbol{G}_{corr,3}$. These $\boldsymbol{G}_{dist,m}$ and $\boldsymbol{G}_{corr,m}$ (*m = 1, 2,* and *3*) are built mimicking the current practice. For $\boldsymbol{G}_{sgm}$ constructed by our new donor gauge selection method, the number of donors for each target site is automatically determined via the SGM algorithm. The same inference method is used for all three methods in the comparison.

### 4.1.3 Estimations of test error and inference accuracy

The test error, $error_{test}$, is computed in the same way as the validation error, $error_{val}$, with equation (29), but using the test set instead of the training set in equations (27-29).

The accuracy of each inferred streamflow at the target gauges associated with the graphs $\boldsymbol{G}_{sgm}$, $\boldsymbol{G}_{dist,m}$ and $\boldsymbol{G}_{corr,m}$ (*m = 1, 2,* and *3*) is evaluated by the Nash–Sutcliffe efficiency coefficient (NSE) (Nash & Sutcliffe, 1970) with the testing data set. The NSE with the testing data set for target site *j* ($NSE_{test_j}$) is computed from the observed ($\mathbf{Q}_{\text{test}_j}$) and the inferred ($\widehat{\mathbf{Q}}_{\text{test}_j}$) streamflow time series as follows,

$$NSE_{test_j} = NSE\left(\mathbf{Q}_{\text{test}_j}, \widehat{\mathbf{Q}}_{\text{test}_j}\right) \qquad (33)$$

### 4.2. Removal of streamflow gauges example

Our method can readily be applied to removal of streamflow gauges (RG) problems. A new RG algorithm is devised to facilitate repeatedly removing the gauge inferred by the remaining gauges with the highest effectiveness. That is, RG first removes gauge $j$ in the network with the highest $NSE_{test_j}$ from the set of $p$ gauges of the network, and then marks the removed gauge as a "target gauge" and each of its neighbors as a "donor gauge"; this process is repeated for the remaining available gauges in the network until all gauges are checked, with the exception that isolated gauges should not be removed. Algorithm 2 given below describes the details of this gauge removal process with the least loss of information. Upon the completion of running the RG algorithm, maxRemRank gives the maximum number of removable gauges, with an accompany of removable-queue that contains all removable gauges in the ascending order of the information loss.

Equation (34) defines a new score, $\boldsymbol{graph\_score_{test}}$, for the graph, based on the NSE, for gauges that can be removed from the hydrometric network. In equation (34), the index $k$ = 'Dist', 'Corr', or 'SGM' represents each of the three approaches, and $\boldsymbol{M_{rem}}$ represents the largest number of removable gauges among all three approaches whose NSEs are greater than or equal to the threshold $\boldsymbol{\delta}$ used.

$$graph\_score_{test,k} = \frac{1}{M_{rem}} \sum_{i=1}^{M_{rem}} NSE_{i,k} \qquad (34)$$

The $graph\_score_{test,k}$ is useful to assess the quality and quantity of the inference of daily streamflow time series for the removable gauges from a given graph $k$. The higher the score, which has a highest value of 1, the better.

**Algorithm 2: Removal of Gauges (RG) Algorithm**

**STEP 0.** RG inputs: $[NSE_{test_1}, \ldots, NSE_{test_q}]$, $\boldsymbol{G}$

**STEP 1.** Mark all isolated nodes in the graph $\boldsymbol{G}$ as unavailable for removal.

**STEP 2.** Initialize the maximum number of removable gauges: maxRemRank=0.

**STEP 3.** Initialize the queue of removable gauges: removable-queue = { }.

**STEP 4.** Sort $[NSE_{test_1}, \ldots, NSE_{test_q}]$ as the current queue of gauges (i.e., current-queue) in the descending order of $NSE_{test}$ values.

**STEP 5.** Check the first item of the current-queue, $NSE_{test_r}$. If the $r$th gauge is not marked as unavailable for removal, do steps 6 and 7; otherwise go to step 8.

**STEP 6.** Add the $r$th gauge into the removable-queue and mark its neighbors on the graph $\boldsymbol{G}$ as unavailable for removal.

**STEP 7.** Update maxRemRank: maxRemRank = maxRemRank + 1.

**STEP 8.** Remove $NSE_{test_r}$ from the current-queue.

**STEP 9.** Repeat from step 5 to step 8 until the current-queue becomes empty

## 5. Study Area and Data Sets

Our application examples use Ohio River basin as the test base for its size, relevance and good quality of long-term historical daily streamflow data. The Ohio River is the third largest river in terms of discharge in the United States. It is the largest tributary of the Mississippi River and accounts for more than 40% of the discharge of the Mississippi River (Benke & Cushing, 2011). The Ohio River is located between the 77° and 89° west longitude and between the 34° and 41° north latitude.

Table 1 lists the National Weather Service Location Identifier (NWSLI) which is used in this study to index each of the 34 gauges, the drainage area of the corresponding sub-basin, and the USGS station identifier. The naturalized daily streamflow data are taken from the National Water Information System (NWIS: National Water Information System), United States Geological Survey (USGS). This data set spans the time period from January 1st, 1951 to December 31st, 1980 with a total of 10958 consecutive days (30 years) of measurements for all of the 34 streamflow gauges. There are no missing streamflow records for any day or gauge over the selected study period.

**Table 1**– List of 34 streamflow gauges over the Ohio River basin

| # | NWSLI | USGS STAID | Drainage Area (Km²) | # | NWSLI | USGS STAID | Drainage Area (Km²) |
|---|-------|-----------|---------------------|---|-------|-----------|---------------------|
| 1 | ALDW2 | 03183500 | 3,533 | 18 | GRYV2 | 03170000 | 777 |
| 2 | ALPI3 | 03275000 | 1,352 | 19 | KINT1 | 03434500 | 1,764 |
| 3 | ATHO1 | 03159500 | 2,442 | 20 | MROI3 | 03326500 | 1,766 |
| 4 | BAKI3 | 03364000 | 4,421 | 21 | NHSO1 | 03118500 | 453 |
| 5 | BELW2 | 03051000 | 1,052 | 22 | NWBI3 | 03360500 | 12,142 |
| 6 | BOOK2 | 03281500 | 1,870 | 23 | PRGO1 | 03219500 | 1,469 |
| 7 | BSNK2 | 03301500 | 3,364 | 24 | PSNW2 | 03069500 | 1,870 |
| 8 | BUCW2 | 03182500 | 1,399 | 25 | SERI3 | 03365500 | 6,063 |
| 9 | CLAI2 | 03379500 | 2,929 | 26 | SLMN6 | 03011020 | 4,165 |
| 10 | CLBK2 | 03307000 | 487 | 27 | SNCP1 | 03032500 | 1,368 |
| 11 | CRWI3 | 03339500 | 1,318 | 28 | STMI2 | 03345500 | 3,926 |
| 12 | CYCK2 | 03283500 | 938 | 29 | STRO1 | 04185000 | 1,062 |
| 13 | CYNK2 | 03252500 | 1,608 | 30 | UPPO1 | 04196500 | 772 |
| 14 | DBVO1 | 03230500 | 1,383 | 31 | VERO1 | 04199500 | 679 |
| 15 | ELRP1 | 03010500 | 1,424 | 32 | WTVO1 | 04193500 | 16,395 |
| 16 | FDYO1 | 04189000 | 896 | 33 | WUNO1 | 03237500 | 1,002 |
| 17 | GAXV2 | 03164000 | 2,929 | 34 | WYNI2 | 03380500 | 1,202 |

Following the procedure described in Section 3.1.2, the dataset was separated into 3 subsets as described in the preceding section. Data between 1951 and 1970 were used for "training" and "validation". The training data set consists of 50% of the data randomly selected from 1951 and 1970. The remaining data over the period of 1951 and 1970 constitute the validation set. The data between 1971 and 1980 are used as the "test" set.

## 6. Results and Discussion

### 6.1 Inference on streamflow

The inferred daily streamflow time series based on the new graphical model donor gauge selection method (i.e., graph $G_{sgm}$) and the conventional distance- and correlation-based methods (i.e., graphs of $G_{dist,m}$ and $G_{corr,m}$ with $m = 1, 2,$ and $3$) are compared. $G_{dist,m}$ and $G_{corr,m}$ with $m = 1, 2,$ and $3$ are constructed as described in Section 4.1.2. For our new method with the SGM algorithm, a Pareto front is obtained and we choose among the candidates on the front the graphs that have similar number of edges as $G_{dist}$ and $G_{corr}$ for fair comparison. Our method was run with default parameters defined in *Algorithm 1*. That is, 30 different values of the regularization parameter $\lambda$ were used for graphs with edges between 10 (very sparse) and 561 (complete graph). Thus, the number of sampling points is (561 - (10-1)) *30 = 16560 (based on Step 2 in Algorithm 1). The SGM algorithm selected 74 out of 16560 (0.45%) distinct graphs with different number of edges as the candidate solutions according to our two multi-objective optimization criteria.

From the distance-based method, 1, 2 and 3 donors result in 24, 43 and 65 edges, respectively, denoted as Dist(24), Dist(43) and Dist(65), respectively; for the correlation-based method, they are Corr(24), Corr(47) and Dist(68). Thus, SGM(25), SGM(47) and SGM(65) are selected from our SGM method to have similar number of edges.

For our method, Figure 2 (a) shows trade-offs between the number of edges and the validation error, $error_{val}$, with the threshold $\Gamma = 0.7$ in equation (27). The black dots represent the dominated solutions in the multiple-optimization space. The three red dots of the non-dominated solutions are from SGM graphs of SGM(25), SGM(47) and SGM(65). The remaining non-dominated solutions (i.e., solutions along the Pareto front) are represented by the green dots. Figure 2 (b) shows the comparison of the mean test error associated each method. The mean test error is calculated based on 500 individual test errors, $error_{test}$, each of which is computed based on the parameters from one random sampling from the training and validation data set but using the same testing data set.

At the top portion of the Pareto front (e.g., green and red dots) in Figure 2 (a), large validation errors occur when the graphs are very sparse. But the error decreases quickly as the number of edges increases until about 44 edges, from where to about 92 edges, the change in error is negligible. At 93 edges there is a noticeable decrease in the validation error. From 93 to about 211 edges the change in validation error is negligible again. The next set of non-dominated solutions is from 211 edges onward with a slight decrease in the validation error where the Pareto front becomes nearly flat and reaches the minimum validation error at 222 edges. For this study region, it appears that a good trade-off between the sparsity and validation error stands at about having 44 or 45 edges; starting with the number of edges around 45, an increase in the number of edges only reduces the error slightly. When the number of edges increases to about 93 or more, the improvement in error reduction becomes almost unnoticeable. Figure 2 (a) also shows that the relationship between the error and the number of edges has an L-shape in which the error approaches almost a constant when the graph grows to have 93 edges. The few "sudden" discontinuities observed in Figure 2 (a) are due to the nature of the error function which includes conditional terms above/below a threshold, $\Gamma$ in equation (27), that might affect continuity of the total validation error when the threshold bound is reached. The complete graph with 561 edges is not in the set of non-dominated solutions, which means that using all of the gauges available in the network to infer the streamflow for the target gauge gives worse results than many of the sparser graphs. This can be attributed to the noisy correlation calculated due to the large noises involved in the data. In fact, Figure 2 (a) shows that using graphs with more than 222 edges is unlikely to reduce the validation error anymore. This result clearly shows that more complexity is not necessarily better.

In Figure 2 (b) the mean test errors for three different levels of sparsity are represented by the three red, green, and magenta bars for the graphs of $\boldsymbol{G}_{sgm}$, $\boldsymbol{G}_{dist}$, and $\boldsymbol{G}_{corr}$, respectively. It shows that the mean test errors are the lowest for the inferred daily streamflow time series using $\boldsymbol{G}_{sgm}$, and are the highest based on $\boldsymbol{G}_{dist}$, with $\boldsymbol{G}_{corr}$ from the pair-wise correlation-based approach lying in the middle.

The statistical significance test of the results shown in Figure 2 (b) was performed as follows: six single tailed t-tests were run using a significance level of 0.05, and a null hypothesis that the mean test error for the SGM graphs is equal to the Dist or Corr graphs; the null hypothesis was rejected in all cases (p-value < 0.0001), and the alternative hypothesis was accepted. That is, the mean error with the testing data set from SGM are all significantly lower than their counterparts from Dist and Corr. In other words, the results obtained using our new donor gauge selection method with the SGM algorithm, as shown in Figure 2 (b), are significantly better than those of using either the least distance-based or the maximum correlation-based approaches.

Figure 2 (c) shows the relationships between the mean test error and the number of training days used for Dist (green), Corr (magenta) and SGM (red), respectively. The length of the training set varies from 45 days to 3650 days (ten years). The fifth point in each curve in Figure 2 (c) corresponds to 730 days (about two years). It shows that two years of training data are almost as good as the full range of 10 years. This result is important because for the case of ungauged basins, it is possible to place a temporary gauge station to collect data for about two years and then use the collected data to train the algorithms presented in this work to infer the streamflow time series for that specific ungauged location in the future provided no dramatic environment change occurs for the study region.

Figure 3 presents graphs obtained using SGM, Dist, and Corr. Only the graphs corresponding to 1-, 2-, and 3-donors per target site for Dist and Corr with their counterparts for SGM are plotted: green edges for Dist, magenta edges for Corr and red edges for SGM. It can be observed that the graphs associated with each of the three approaches are different albeit some features in their graphic structures are similar.

From the comparisons shown in Figure 2 (b) and the statistical significance testing results, it is clear that the new SGM method is the best of the three. This is because our new method with the SGM algorithm accounts for the dependence structure in the entire streamflow network based on the concept of conditional independence and employs the Glasso method to effectively extract such dependence structure by making the precision matrix sparse. Our results demonstrate that a good use of the conditional independence structure of the underlying streamflow network (i.e., use sparse precision matrix) is important and it outperforms the widely used pair-wise correlation-based method (i.e., Corr) which only directly uses the local correlation information. Furthermore, the conditional independence structure embedded in the graph $\boldsymbol{G}_{sgm}$ guarantees that all multiple donor gauges identified for each target gauge are not linearly dependent. If they were, some of these gauges would be conditionally independent to the target gauge given the other identified donor gauges, and therefore there would be no direct link(s) between such gauge(s) and the target gauge in our $\boldsymbol{G}_{sgm}$. However, for the Dist and Corr methods, the linear dependence issue is likely to occur among their identified multiple donors. Our results also confirm that in comparison with the distance-based method (Dist), the correlation-based method (Corr) performs much better as reported in the literature (e.g., Archfield & Vogel, 2010; Ergen & Kentel, 2016; He et al., 2011). Again, the different results are simply due to the different donor gauge selection methods as the inference method used is the same in all of the scenarios.

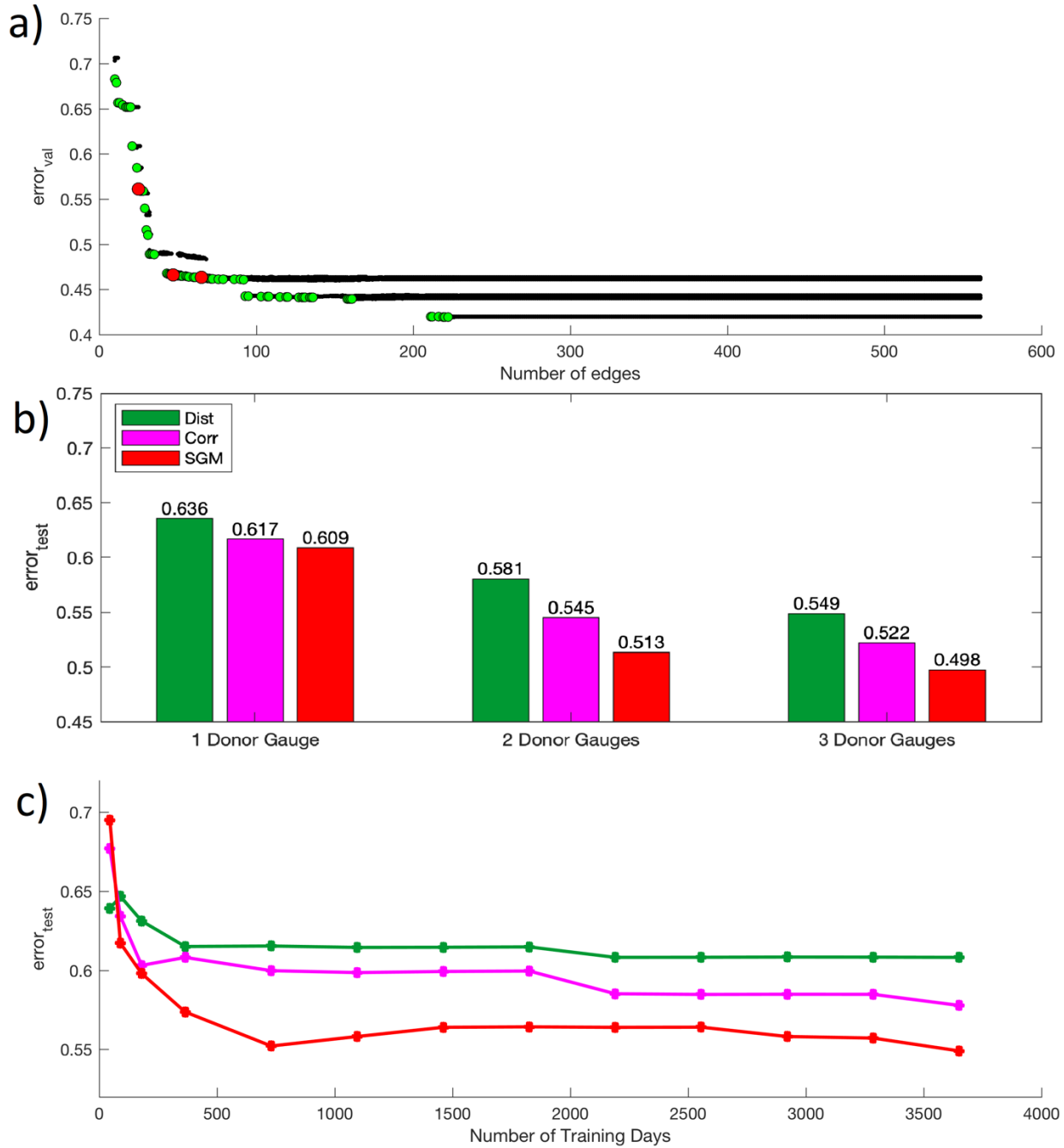

**Figure 2.** Result with the Ohio River Basin dataset. The training set is composed by a random selection of daily streamflow records between 1951 and 1970, while the validation set is composed by the remaining 50% dataset for the same time span. The test data set is composed by the most recent 10 years of data (i.e. 1971-1980). **(a)** Validation error from the multi-objective optimization procedure of the SGM algorithm, between the observed and inferred daily streamflow time series versus the number of edges in the underlying graph. The black dots represent sub-optimal (dominated) solutions. The green dots represent the set of non-dominated (optimal) solutions. The red dots represent the graphs SGM(25), SGM(47) and SGM(65) with 25, 47 and 65 edges, respectively, chosen from the set of non-dominated solutions. **(b)** Comparison of the mean test error among our new method (SGM algorithm) and the other two donor gauge selection methods of the least distance (Dist) and maximum correlation (Corr) for 1, 2 and 3 donor gauges, respectively. From left to right, Dist(24) and Corr(24) with one donor gauge and their counterpart of SGM(25); Dist(43), and Corr(47) with two donor gauges and their counterpart of SGM(47); and Dist(65) and Corr(68) with three donor gauges and their counterpart of SGM(65). **(c)** Comparison of the mean test error as a function of the number of training days. Each point is calculated by averaging the values of the test error for 1-, 2-, and 3-donor gauges or their equivalent counterparts in the SGM case using a given number of training days. The series for Dist, Corr and SGM algorithm are depicted in green, magenta and red, respectively. The size of the training sets used is 45, 90, 180, 365, 730, 1095, 1460, 1825, 2190, 2555, 2920, 3285, and 3650 days, respectively.

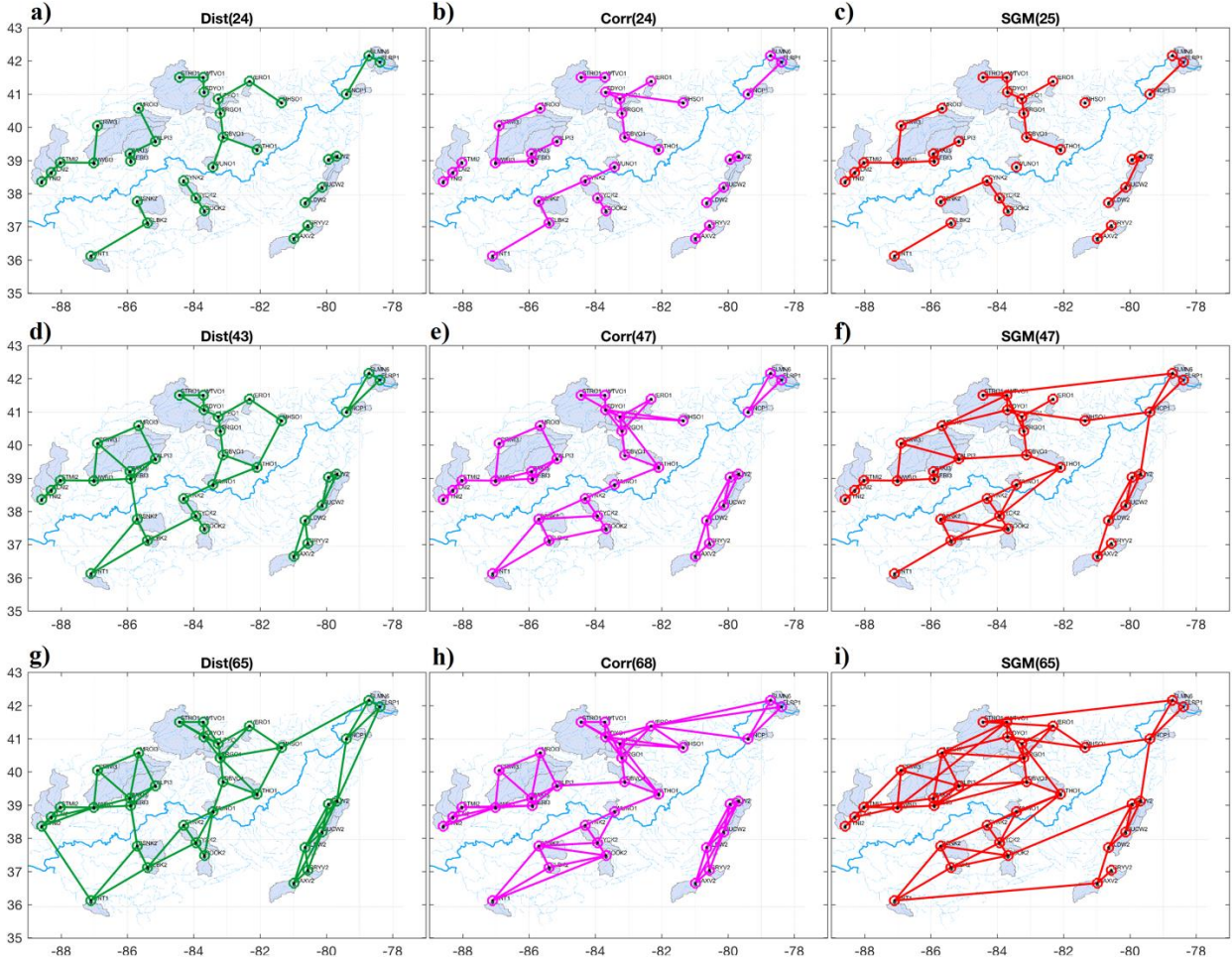

**Figure 3** – Comparison of the graphs generated by Dist on the left, the Corr in the center and SGM on the right, for 1 (on the top), 2 (in the middle) and 3 (at the bottom) donor gauges, respectively, for both Dist and Corr. The number of edges in the graphs for the 'Dist' and 'Corr' methods is fixed albeit depending on the number of donor gauges used to build it, as opposed to the graphs from the SGM algorithm, where the number of edges is selected from the set of non-dominated solutions to approximately match the sparsity of the 'Dist' and 'Corr' graphs for 1-, 2- and 3-donor gauges, respectively. **(a)** Dist(24) with a single donor gauge. **(b)** Corr(24) with a single donor gauge, **(c)** SGM(25) with 25 edges. **(d)** Dist(43) with two donor gauges. **(e)** Corr(47) with two donor gauges. **(f)** SGM(47) with 47 edges. **(g)** Dist(65) with three donor gauges. **(h)** Corr(68) with three donor gauges. **(i)** SGM(65) with 65 edges.

## 6.2 Removal of gauges with least loss of information

We demonstrate how easily our new method can be applied to gauge removal problems. One can simply take each of the nine graphs plotted in Figure 3 as the input separately to the Removal of Gauges (RG) algorithm given in Section 4.2. For this study the threshold value of $\delta$ used for $M_{rem}$ in equation (34) was set to 0.7. While the results of RG algorithm indicate that there are $8 - 16$ gauges that are potentially removable, but only about $7 - 8$ gauges of them can be inferred with an NSE higher than the set threshold value δ of 0.7 if removed. It is important to note that the RG algorithm (i.e., Algorithm 2) does not use the δ. It simply tries to remove gauges until there are no more available gauges to be removed, i.e., until the current-queue in Algorithm 2 becomes empty. On the other hand, the threshold value δ is used to assess the number of gauges that can be removed with given confidence, i.e., when their corresponding NSEs are greater than or equal to δ. This is why there are 8-16 gauges can be removed using the RG algorithm, but only a sub-set of them (i.e., 7-8 gauges) satisfies NSE > 0.7.

Figure 4 shows the results of comparison where removable gauges are represented by color-coded solid circles indicating their corresponding inference accuracy measured by an NSE value. NSE ≥0.9 is drawn in blue; $0.8 \leq NSE < 0.9$ in green; $0.7 \leq NSE < 0.8$ in yellow; $0.6 \leq NSE < 0.7$ in orange; and NSE< 0.6 in red. From Figure 4, it can be clearly observed that, in general, more gauges are removable based on the SGM approach. Furthermore, for removing the same gauges by all three methods, the information lost is the lowest by the SGM method as there are more combined blue, green and yellow solid circles present in Figure 4 for the SGM method.

Figure 5 (a) shows the average graph scores based on the 500 simulations with random sampling from the training and validation data sets but using the same testing data set for the single, two, and three-donor cases, respectively. Each of the graph scores is calculated using equation (34). It can be seen that the average graph scores are higher for the SGM approach than those of the other two approaches in all three cases (i.e., the single, two, and three donors). The SGM approach significantly outperforms the other two methods in the case with two donors. Figure 5 (b) shows the mean graph score calculated by averaging the three average graph scores for each method. It can be seen that the mean graph score of 0.811 for the SGM method is the highest, the Corr method ranked second with 0.762, and the Dist method the lowest at 0.738. Figure 5 (c) shows a related but slightly different measure from that used for Figure 5 (a) in assessing the quality of the inference results for the estimated streamflow time series. In Figure 5 (c), the mean is taken from the eight removable gauges with the highest NSE for each of the donor scenarios. The SGM, again, has the highest mean among the top eight removable gauges. Similar to Figure 5 (b), Figure 5 (d) shows the mean NSE presented in Figure 5 (c) for each of the three methods. Clearly, the mean NSE for the top eight removable gauges for the SGM method is the highest. To further support the observed differences, single tailed t-tests are conducted in a pair-wise fashion, that is, each result is tested against results from each different method with compatible number of edges. The results show that all the differences as observed in Figures 5 (a), (b), (c) and (d) are all statistically significant at a significance level of 0.05.

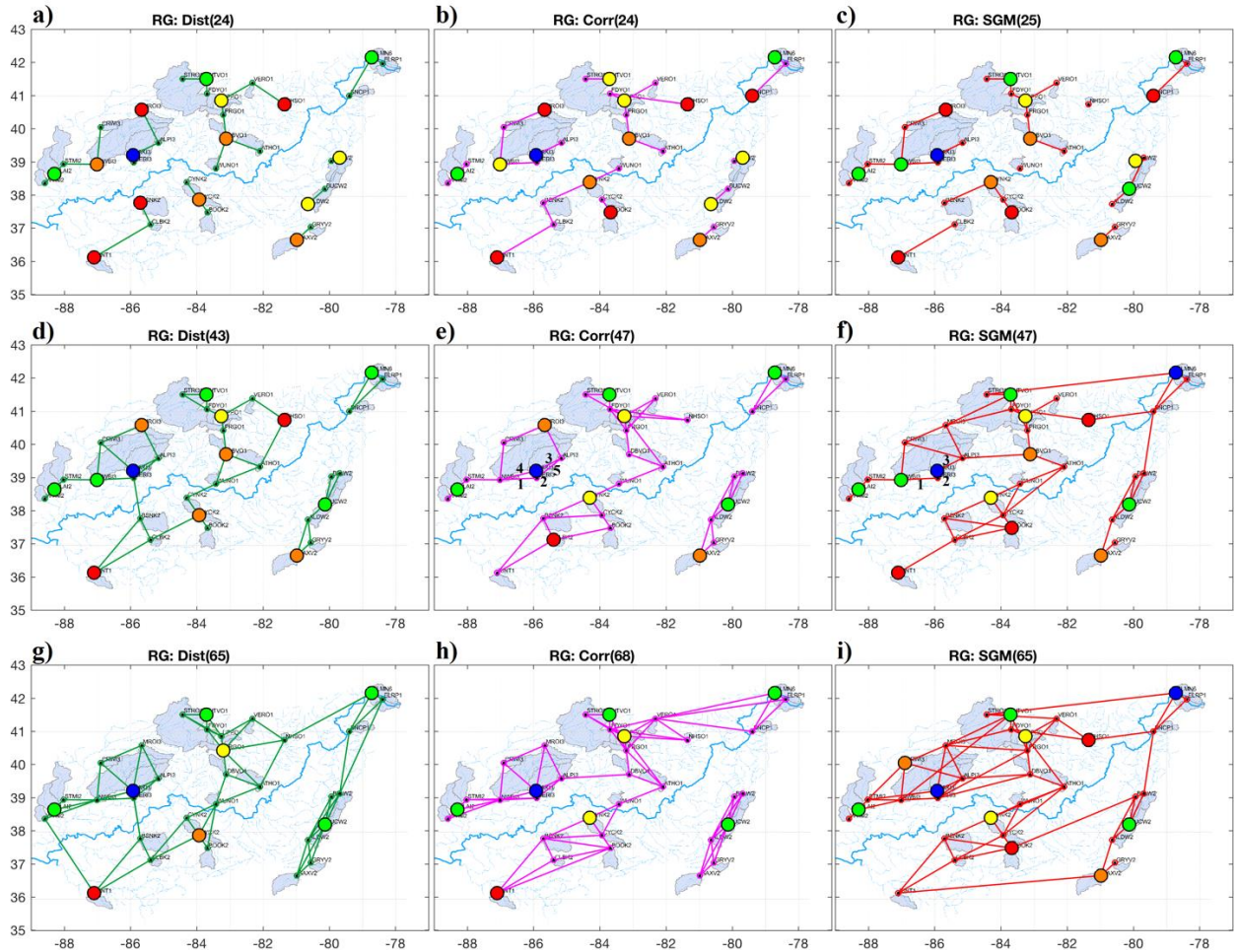

**Figure 4** – Comparison of the observed and inferred daily streamflow time series, in the *test* set (records between 1971 and 1980), for removable gauges estimated by the Removal of Gauges (RG) algorithm using the graphs in Figure 3 as inputs, with Dist on the left, Corr in the middle, and SGM on the right, for donor gauges of 1 (on the top), 2 (in the middle) and 3 (at the bottom), respectively for Dist and Corr methods. Note that for the SGM case, the number of donor gauges are not fixed but automatically determined. The target gauges chosen by the RG algorithm are highlighted in blue for Nash Sutcliffe efficiency (NSE) ≥ 0.9, in green for 0.8 ≤ **NSE** < 0.9, in yellow for 0.7 ≤ **NSE** < 0.8, in orange for 0.6 ≤ **NSE** < 0.7, and in red for NSE < 0.6. **(a)** Dist(24) **(b)** Corr(24) **(c)** SGM(25) **(d)** Dist(43) **(e)** Corr(47) in which five edges are marked in the plot as: 1 (NWBI3-SERI3), 2 (BAKI3-SERI3), 3 (ALPI3-BAKI3), 4 (NWBI3-BAKI3), and 5 (ALPI3-SERI3) **(f)** SGM(47) in which three edges are marked in the plot as: 1 (NWBI3-SERI3), 2 (BAKI3-SERI3), and 3 (ALPI3-BAKI3). **(g)** Dist(65) **(h)** Corr(68) **(i)** SGM(65).

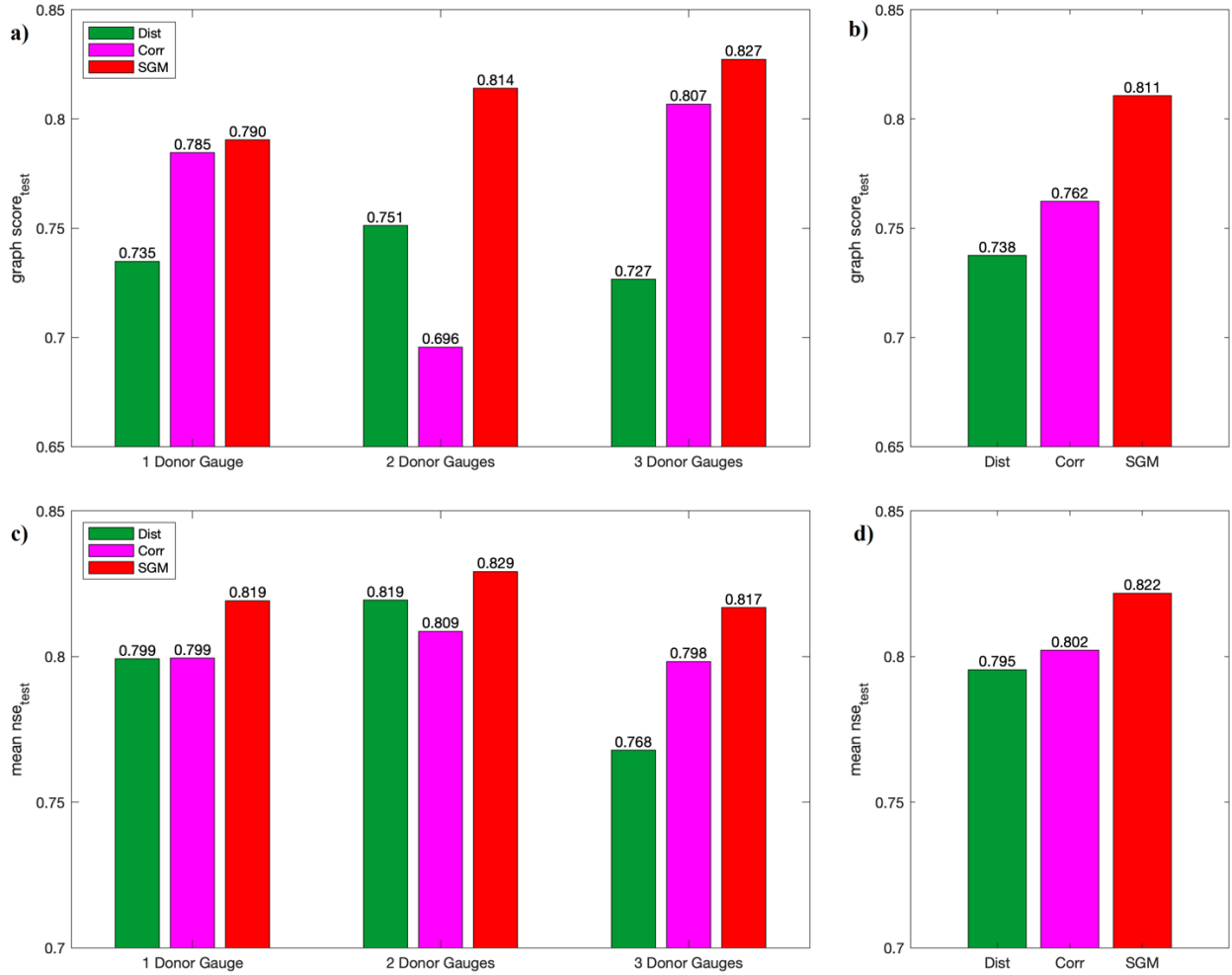

**Figure 5** - Comparison of the mean inference accuracy on the removable gauges with the RG algorithm applied to graphs corresponding to methods of SGM, Dist and Corr. **(a)** Averaged graph score calculated by equation (34) for each of the individual graphs based on 500 simulations with random samples from the training and validation data sets but with the same testing data set (i.e., $1971 - 1980$). **(b)** The mean graph score averaged over the 1-, 2- and 3-donors based on values shown in (a) for each method. **(c)** Averaged NSEs of the eight removable gauges with the highest NSEs based on the 500 simulations. **(d)** The mean NSE averaged over the 1-, 2- and 3-donors based on values shown in **(c)** for each method.

## 6.3 Insights provided by SGM

In general, the pair-wise correlation-based approach (Corr) is again more accurate than the distance-based approach (Dist), since the former requires more and better data to establish the correlations. From the correlation perspective, the fundamental difference between the SGM method and the widely used Corr method is that the Corr method uses the pair-wise correlation to determine the edges between the gauges, whereas the new SGM method takes advantage of the conditional independence structure among all gauges in our graphical model. Due to the pair-wise correlation nature of the Corr method, any gauge may have a decent pair-wise correlation with a relatively large number of other gauges where some of these pair-wise correlations are actually redundant. In contrast, the new SGM method is able to reveal the conditional independence structure among the gauges within the entire network, and therefore only identifies those gauges that can make unique (i.e., not redundant) contributions to the estimation for target gauge. This desirable characteristic of the SGM method enables to capture the cleaner and more accurate dependence structure for each individual gauge in the context of the underlying gauge network. This, in turn, also leads to the identification of more gauges that can be inferred without a significant loss in accuracy. Our results in Figures 4 and 5 have shown that indeed the accuracy of the inferred streamflow time series is improved and that the number of potentially removable gauges is also increased compared to the Corr approach.

One clear example of the difference between the Corr and the SGM methods is manifested by the relationship identified between the sites ALPI3, BAKI3, NWBI3 and SERI3 shown in Figures 4 (e) and 4 (f). BAKI3, with a catchment area of 4421 $Km^2$, is a sub-basin of SERI3 with a catchment area of 6063 $Km^2$ along the main channel. Therefore, the catchment area of BAKI3 accounts for 73% of the catchment area of SERI3 and the correlation between them is the highest among the sites considered in the study area. The edge between them is present in all of the nine graphs shown in Figure 4. The sites BAKI3 and SERI3 are also highly correlated to the sites ALPI3 and NWBI3. Figure 4 (e) shows the graph for Corr(47), with five edges: 1 (NWBI3-SERI3), 2 (BAKI3-SERI3), 3 (ALPI3-BAKI3), 4 (NWBI3-BAKI3), and 5 (ALPI3-SERI3). Figure 4 (f) shows the graph for SGM(47) with only three edges which are the subset of the edges with Corr(47), by having edges NWBI3-BAKI3 and ALPI3-SERI3 dropped. It is safe for SGM(47) to drop these two edges as NWBI3 is conditionally independent to BAKI3 given SERI3, and ALPI3 is conditionally independent of SERI3 given BAKI3. Figures 5 (a) and 5 (c) show that among the nine graphs shown in Figure 4 the graph with the best trade-off between model complexity and accuracy is SGM(47).

Figure 6 shows the comparison between the observed and inferred daily streamflow time series based on the testing set for the eight streamflow gauges with the highest NSE, when SGM(47), shown in Figure 4 (f), is chosen as the underlying graphical model.

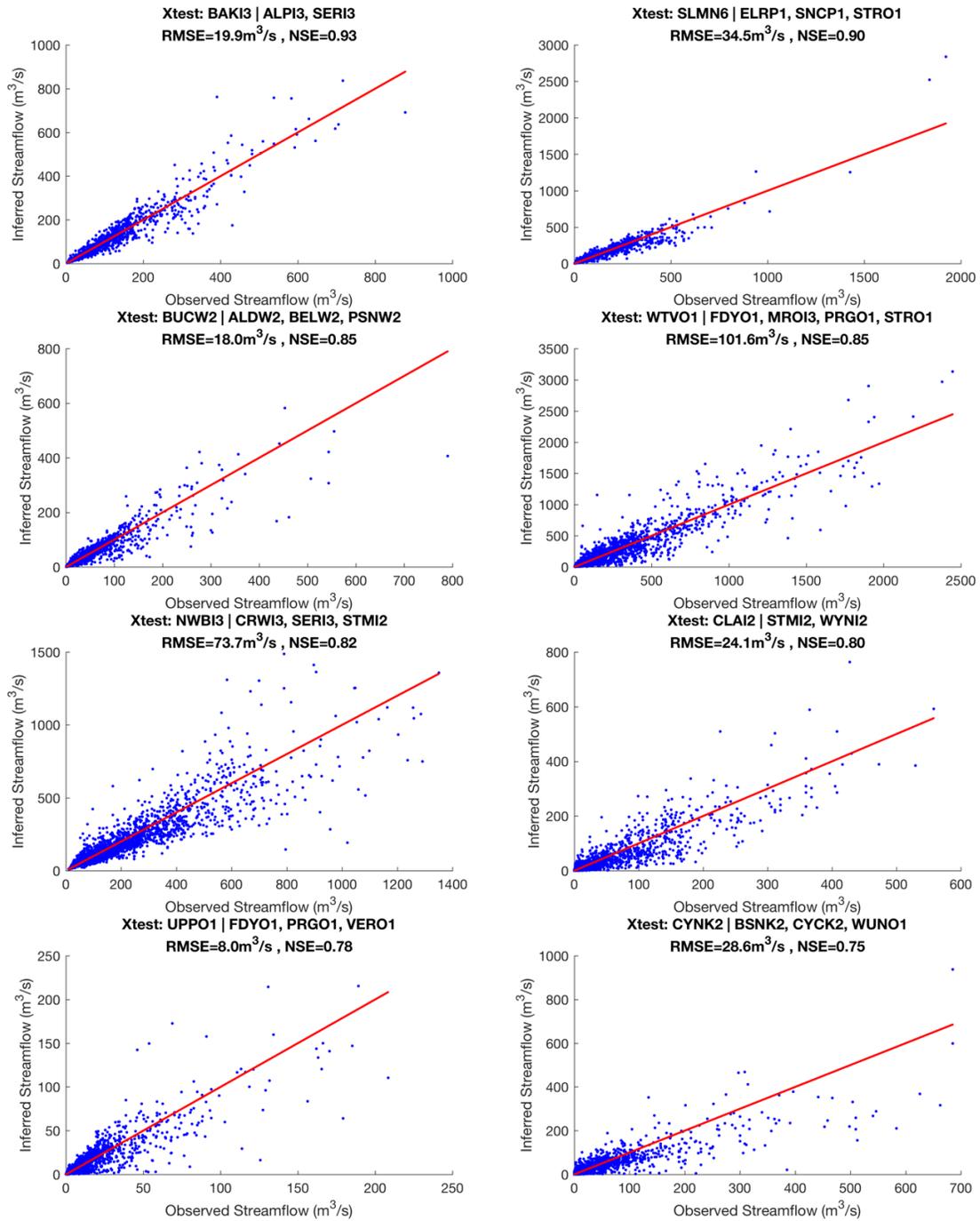

**Figure 6.** Scatter plots between the observed and inferred daily streamflow time series over the test period of 1971-1980 (i.e., test data set). Each plot represents one of the eight gauges with the highest NSE values among the removable gauges shown in Figure 4 (f). The RG algorithm is used upon SGM(47) to identify the gauges to be removed. The MLR with equations (31) and (32) is used to infer the daily streamflow shown in the plots. The root mean squared error (*RMSE*) and the *NSE* are shown for each gauge over the inferred 10 years of 1971-1980. At the top of each plot,

the name of the removed gauge is indicated on the left side of the divide line "|", and the names of gauges used to infer the streamflow of the removed gauge are indicated on the right side of the divide line "|".

For each of the 34 gauges, their corresponding watersheds were delineated using a Geographical Information System (GIS) to facilitate our understanding of the identified connections and isolated gauges based on the SGM method. Figure 7 shows the elevation (NED: National Elevation Dataset), slope (derived from elevation data), soil type (Hybrid STATSGO/FAO Soil Texture) and land cover (MRLC: Multi-Resolution Land Characteristics Consortium) along with the selected non-dominated graph SGM(25).

On SGM(25), NHSO1 and WUNO1 are two isolated sites. Isolated sites should be maintained as much as possible to avoid loss of important regional information. Less sparse graphs, such as SGM(47) and SGM(65), have some marginal benefits from having some edges to those sites. NHSO1 has a significantly different land use comparing to other watersheds in the study region. For NHSO1, more than 50% of its drainage area is developed while others have less than 20%. Thus, the hydrological response of this watershed to precipitation events is very different from other watersheds. In the case of WUNO1, its isolation in SGM(25) appears to be related to a combination of its geographic location, different land use from its neighboring watersheds, and its proximity to the main channel of the Ohio River. This last factor seems to be a natural separator for it. There are no edges crossing the Ohio River on the selected sparse graph, SMG(25), as shown in Figure 3 (c).

The factors that impact the connections (i.e., conditional dependence) between gauges are complex and it is the integrated effect (e.g., the streamflow in this case) that determines the conditional dependence between the gauges. The prime factors that contribute to the generation of streamflow in the study area seem to be the elevation, slope and catchment area. There is a relatively high correlation between the specific discharge (i.e., streamflow divided by the catchment area) and the elevation (0.79), and between the specific discharge and the slope (0.76). The land cover also plays an important role, as the edges in SGM(25) are usually present between sites with the same land cover class as shown in Figure 7 (d).

Results here have demonstrated again that it can be difficult to justify the use of relatively simple and explicit functions to relate streamflow to different factors such as land cover, slope, soil type, drainage size in identifying their connections for complicated situations like this study case. This point has previously been demonstrated in the literature (e.g., Parada and Liang, 2010). On the other hand, these factors may shed lights on why certain links exist while others do not. For example, the land cover types, elevation, and slopes appear to play more important roles than the soil type in this study region. It is worth pointing out that gauges are sometimes connected even if the correlations between them are not very high. They are connected simply because there are no other available gauges nearby with acceptably higher (conditional) dependence. In summary, the chosen graph SGM(25) does not have any edges crossing the Ohio River; there exist two gauges isolated from the rest, those gauges are geographically far from other gauges and one of them has a significantly different land use category distribution with more than 50% of its area being developed. Most of the area of the Ohio River basin belongs to the same soil type category and therefore, the soil type does not appear to contribute to the identification of the hydrologic similarity between sub-basins in this study case. On the other hand, most of the edges on the selected underlying graph SGM(25) are between watersheds with the same land use category.

These results suggest that in the Ohio River basin, the land use is an important factor for the hydrologic similarity among the sub-basins.

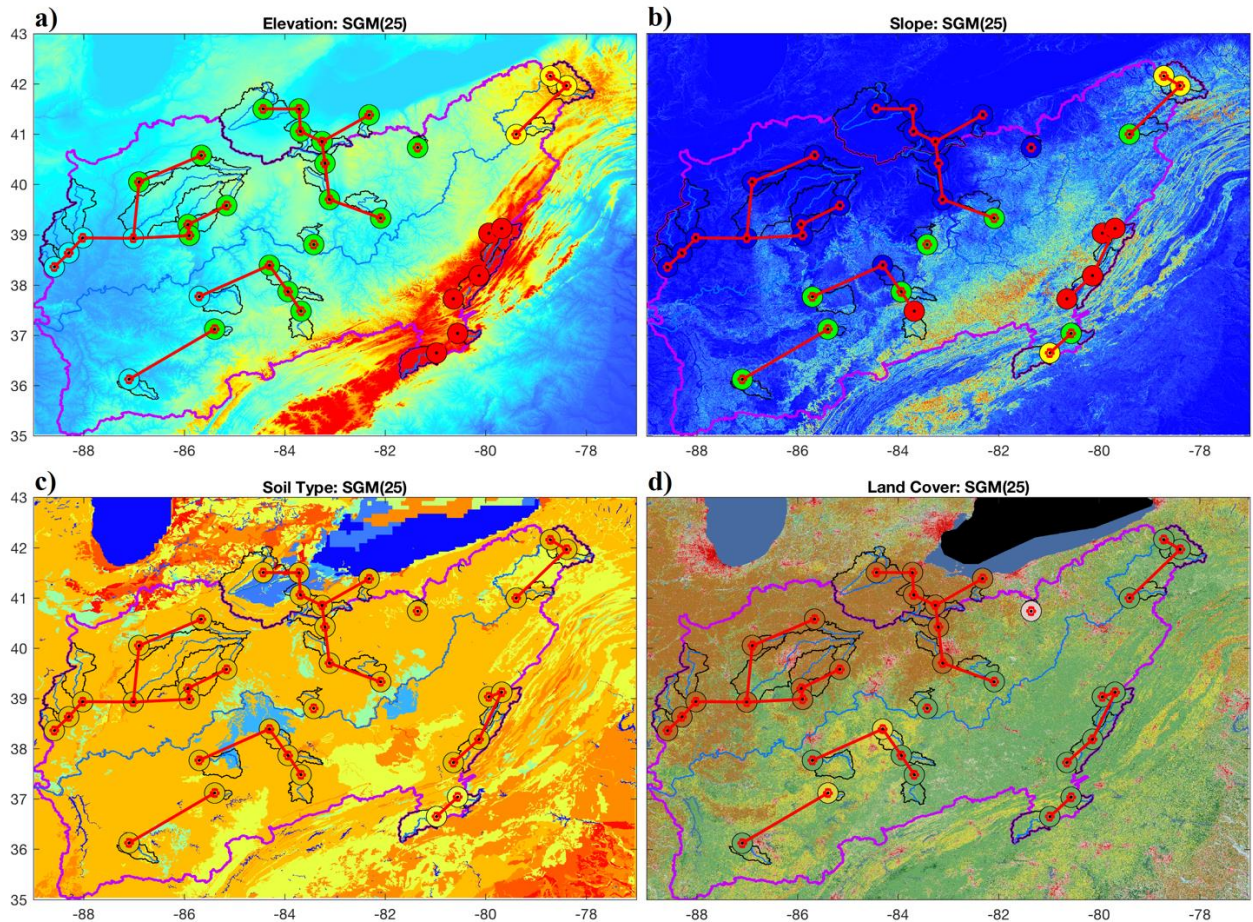

**Figure 7.** Spatial distributions of the elevation, slope, soil type, and land cover over the study region structured with graph SGM(25). **(a)** Elevation map showing a cluster of 4 categories indicated by a filled circle of cyan, green, yellow and red respectively. These different colors represent, respectively, "very low", "low", "high" and "very high" elevations based on the mean elevation of their corresponding watersheds. **(b)** Slope map showing a cluster of 4 categories indicated by a filled circle of blue, green, yellow and red respectively. The colors represent, respectively, "very low", "low", "high" and "very high" slope based on the mean slope of their corresponding watersheds. **(c)** Soil type map showing a cluster of 2 categories indicated by a filled circle of dark yellow and light yellow, respectively. These two colors represent, respectively, the "silt loam" and "loam" soil types. **(d)** Land cover map showing a cluster of 4 categories indicated by a filled circle of pink, green, yellow and brown, respectively. These different colors represent,

respectively, the "developed, open space", "deciduous forest", "pasture/hay" and "cultivated crops" land cover types.

## 7. Conclusions

We present a novel donor gauge selection method that utilizes the global conditional independence structure of a streamflow network based on graphical Markov modeling. An undirected graphical model known as Gaussian Graphical model is first built by mapping all gauges in the network into vertices, and each pair of vertices is linked by an edge if the streamflow time series recorded at their corresponding gauges are conditionally dependent. This conditional independence structure is extracted from the inverse of the data covariance matrix, i.e., the precision matrix, since the covariance matrix does not present such conditional independence information. Because of the high level of noises included in the streamflow data, the precision matrix directly computed from the covariance matrix is not sparse, typically leading to a dense streamflow network graph. The Graphical Lasso (Glasso) method is employed to make the graph sparse so that the conditional independence structure defined in the streamflow network is sharpened. At the core of our method is the *Selection of Graph Model* (SGM) algorithm that we developed to address the specific characteristics of the donor gauge selection problem: First, there must be some threshold level of the conditional dependence determined by parameter $\tau$ in equation (22) and below which a link is not warranted. Also, an L1 norm regularization parameter $\lambda$ in equation (23) is needed in the Glasso method to trim the graph sparse while using a different $\lambda$ results in a different graph. Second, a suitable set of donor gauges taken from candidate graphs should be the set where its streamflow data can be used in inferring those at the target gauge site within acceptable errors. By identifying complexity and error associated with each graph, our SGM algorithm formulates an iterative multi-objective optimization procedure: These two parameters, $\lambda$ and $\tau$ are determined by minimizing both the complexity, i.e., the number of edges of the underlying graph, and the error with the validation data set to achieve a balance between the sparsity/connectivity and accuracy for each graph.

We have also developed a new algorithm, called the Removal of Gauges (RG) algorithm, for use in streamflow gauge removal application, should there be such a need to discontinue gauges due to budget constraints. The RG algorithm identifies a set of gauges for removal from an existing hydrometric streamflow network with the least loss of information using the SGM algorithm.

In this study, the strengths and effectiveness of our new donor gauge selection method are illustrated through two types of applications, record extension and gauge removal, by using daily streamflow data from a hydrometric network of 34 gauges from the Ohio River Basin between 1 January 1950 and 31 December 1980. We have demonstrated that the selected donor gauges based on the graphs generated by our SGM algorithm lead to more accurate estimates of the daily streamflow time series than the donor gauges selected from the conventional distance-based (Dist) method and the pair-wise correlation-based (Corr) method. In addition, we have illustrated that the graph with 47 edges selected based on the SGM algorithm has a good trade-off/balance between the network sparsity and the estimation error for the Ohio River basin. The graphs obtained also shed lights on why different gauges are conditionally dependent. For the record extension/gap-filling application, the inferred daily streamflow at the target gauges based on the selected donor gauges from our SGM algorithm are compared with those from the other two methods. The comparisons are carried out for three cases which correspond to using 1-, 2-, and 3-donors. In all of these three cases, the inferred streamflow based on our new donor gauge selection method outperforms the others as shown in Figure 2 (b). It is worth pointing out that once the donor gauges are identified, one may apply any inference method, including the advanced machine learning methods, to estimate the streamflow for the target gauges. In this study, we applied a regression-based inference method described in Section 2 because of its simplicity and convenience.

By applying our RG algorithm, we have also demonstrated (e.g. Figure 4 (f) and Figure 5 (c)) that eight out of 34 (24%) gauges can potentially be removed (NSE $\geq$ 0.70), from which a group of six gauges (18%) can be inferred with relatively high accuracy (NSE $\geq$ 0.8) using the donor gauges identified by the SGM algorithm. In contrast, only seven gauges (21%) can be removed with NSE $\geq$ 0.70 for either the least distance (Dist) method or maximum pair-wise correlation (Corr) methods, in the two donor gauge cases. Furthermore, the averaged graph score (i.e., equation (34)) for Dist and Corr methods for the two donor gauge cases are much lower than that from our new method as observed in Figure 5 (a), indicating that the sum of the NSEs are much lower if the same number of gauges are removed. In the gauge removal comparison study, we again used the same regression-based inference method for all three different methods. Depending on the number of gauges needed for removal, a balance between the inference accuracy and the gauge removal numbers can be achieved as demonstrated. In general, the sparser the graphs are, the more gauges can be removed.

Our study also demonstrates that the complete graph (i.e., with 561 edges) is not included in the set of non-dominated solutions of our graphical modeling, indicating that having more donor gauges does not achieve optimum results due to significantly more noises, and therefore, inconsistency, introduced by the data and the inclusion of redundant information. Therefore, not only can a suitable sparse graph identify more desirable donor gauges and thus achieve better inferring results, through finding the most essential correlations from a global point of view, but also it is more practical because the sparse graph gives a small but most relevant number of donor gauges for inferring the streamflow for the target gauges and requires a fewer observations to establish the relationship through the data training process. Furthermore, a graph with a fewer edges can reduce overfitting.

Sensitivity regarding the length of daily data required for achieving a stable SGM graph was investigated. Our results (see Figure 2 (c)) show that a length of about 2-year daily data is needed. Such a short data length requirement has a good implication for potential applications to ungauged basins. For example, one can install a temporary streamflow gauge to collect data for about two years and then use the collected data in combination with data from other existing gauges in the network to obtain the SGM graph. Then, daily streamflow time series for the ungauged basin can be inferred using the methodology described in this study as long as there is no dramatic environment change over the study region. This potential application will be tested and its results will be compared to other methods for inferring the daily streamflow data at the ungauged basins in the future.

There are some requirements that have to be met in applying our method. First, a historical record of two years or more is required to characterize the relationships between the target and donor gauges. Second, the probability distribution of the daily streamflow should be well approximated by a log-normal distribution. This, however, is not an issue even if the log-normal assumption does not hold as it can readily be amended using a common distribution transformation method. It is also to be noted that even though the streamflow inference was performed in this study using an ordinary least squares MLR approach because of its simplicity and convenience, other inference approaches can also be used once the sets of donor gauges for each target gauge are identified by our new SGM algorithm.

The most computationally intensive part in our new method lies at the execution of the SGM algorithm, for it calls the Graphical Lasso method multiple times to find the optimal combination of the regularization parameter $\lambda$ and truncation parameter $\tau$. Even so, in our Ohio River Basin network application, it only took 18 minutes to finish identifying the donor gauge selection (e.g., Figure 2 (a)) using a 2016 MacBook Pro. It is possible to further cutting down the computational efforts. In this work, we performed a thorough search for the regularization parameter between 0 and 1, but it was later found that its best range was between 0.01 and 0.1, which means that significant speedup could be achieved if one just search within this bound instead of values over a wider bound. Also, we performed in this study using an almost exhaustive search for the truncation parameter by going from a very sparse graph with only 10 edges to a full graph with 561 edges (for a graph of 34 nodes) as shown in Figure 2 (a). In fact, sparser graphs may achieve better overall results as discussed above; so, in the future, the computational time can be reduced.

In this work, only contemporaneous daily streamflow records are considered. The methods explained here can be adapted to include lagged records for a finite set of days. However, for the sake of simplicity such approach was not followed. Related work (Farmer, 2016; Skøien & Blöschl, 2007) found only marginal improvements when considering streamflow travel times in geostatistical analysis.

**Acknowledgments and Data**

This work was partially supported by the William Kepler Whiteford Professorship from the University of Pittsburgh.

For this work, German A. Villalba implemented the research ideas, designed algorithms, performed experiments, conducted analysis, and co-wrote the manuscript. Xu Liang conceived the research ideas, designed experiments, supervised the investigation, synthesized the results, and co-wrote and finalized the manuscript. Yao Liang conceived the research ideas and co-wrote the manuscript.

We would like to thank the scientists who worked at the U.S. Geological Survey for their streamflow data (https://waterdata.usgs.gov/nwis) and digital elevation model (https://viewer.nationalmap.gov/basic/?basemap=b1&category=ned), the National Center for Atmospheric Research for the soil type and vegetation data (https://ral.ucar.edu/solutions/products/noah-multiparameterization-land-surface-model-noah-mp-lsm), and the Multi-Resolution Land Characteristic Consortium (https://www.mrlc.gov/) for their land-cover data. All of the data used in this study can be freely obtained from their respective websites as they are all publicly available. The Matlab codes for running our new algorithms and the datasets used to generate the tables and figures in this paper can be accessed from DOI: 10.5281/zenodo.3634206.